\documentclass[onefignum,onetabnum]{siamart250211}
\usepackage{amssymb}
\usepackage{graphicx}
\usepackage{xspace}
\usepackage{multirow}
\usepackage{bbm}
\usepackage{mydef}
\usepackage{remarks}
\usepackage{algorithm}
\usepackage{algpseudocode}
\usepackage[numbers,sort]{natbib}
\usepackage{graphicx}
\usepackage{subcaption}
\usepackage{enumitem}
\usepackage{multirow}
\usepackage{adjustbox}
\usepackage{pdfpages}
\usepackage{booktabs}
\usepackage{tikz}
\usepackage{cancel}

\usepackage{siunitx}
\newcommand{\mynum}[1]{{\num[%
  round-precision=4,
  exponent-mode=threshold,
  exponent-thresholds=-3:3,
  retain-zero-exponent=false]{#1}}}

\let\cite=\citet

\begin{document}

\title{Revisiting the Slip Boundary Condition: Surface Roughness as a Hidden Tuning Parameter%
  \thanks{Draft version of \today}}

\author{
  Matthias~Maier\footnotemark[2]
  \and
  Peter~Munch\footnotemark[3]
  \and
  Murtazo~Nazarov\footnotemark[4]}

\maketitle

\renewcommand{\thefootnote}{\fnsymbol{footnote}}
\footnotetext[2]{%
  Department of Mathematics, Texas A\&M University, 3368 TAMU, College
  Station, TX 77843, USA. \url{maier@tamu.edu}}
\footnotetext[3]{%
  Institute of Mathematics, Technical University of Berlin,
  Straße des 17. Juni 136, 10623 Berlin, Germany.
  \url{muench@math.tu-berlin.de}}
\footnotetext[4]{%
  Division of Scientific Computing, Department of Information Technology,
  Uppsala University, Lägerhyddsvägen 1, 752 37 Uppsala, Sweden.
  \url{murtazo.nazarov@uu.se}}
\renewcommand{\thefootnote}{\arabic{footnote}}

\tableofcontents

\begin{abstract}
  In this paper, we investigate the effect of boundary surface roughness on
  numerical simulations of incompressible fluid flow past a cylinder in two
  and three spatial dimensions furnished with slip boundary conditions. The
  governing equations are approximated using a continuous finite element
  method, stabilized with a Galerkin least-squares approach.

  Through a series of numerical experiments, we demonstrate that: $(i)$ the
  introduction of surface roughness through numerical discretization error,
  or mesh distortion, makes the potential flow solution unstable; $(ii)$
  when numerical surface roughness and mesh distortion are minimized by
  using high-order isoparametric geometry mappings, a stable potential flow
  is obtained in both two and three dimensions; $(iii)$ numerical surface
  roughness, mesh distortion and refinement level can be used as control
  parameters to manipulate drag and lift forces resulting in numerical
  values spanning more than an order of magnitude.

  Our results cast some doubt on the predictive capability of the slip
  boundary condition for wall modeling in turbulent simulations of
  incompressible flow.
\end{abstract}



\pagestyle{myheadings}
\thispagestyle{plain}
\markboth{Maier, Munch, Nazarov}{Surface roughness as a hidden tuning parameter}


\section{Introduction}
We investigate the effect of boundary surface roughness on numerical
simulation of incompressible fluid flow past a cylinder in two and three
spatial dimensions equipped with slip boundary conditions. We focus on the
classical bluff body problem, specifically, flow past a cylinder in both
two and three spatial dimensions. Our main objective is to investigate how
the surface roughness of the cylinder---introduced by numerical
discretization error and mesh distortion---influences the qualitative
nature and quantitative behavior of the simulated flow. Understanding such
an influence on the qualitative and quantitative behavior of the flow is of
particular importance as using slip boundary conditions for wall modeling
in turbulent simulations has been used extensively~\citep{Hoffman_2010,
Hoffman_2011, Min_Kim_2004, Bose_2014, Jansson_2022, Jansson_2025}.

\subsection{Incompressible fluid flow, the invsicid limit, and boundary
conditions}

Incompressible fluid flow is modeled mathematically by the following
Navier-Stokes equations: Let $\Omega$ be an open domain in $\mR^d$,
$d=2,3$, and let $T>0$ be the final time. We seek a velocity field
$\bu:=(u_1, \ldots, u_d)$ and a pressure $p$ solving the partial
differential equations
\begin{equation}\label{eq:ns}
  \begin{cases}
    \begin{aligned}
      \p_t \bu + \bu \SCAL \GRAD \bu + \GRAD p - 2\nu \DIV
      \bvarepsilon(\bu) &= 0,
      \\
      \DIV \bu &= 0,
    \end{aligned}
  \end{cases}
\end{equation}
for every  $(\bx,t) \in \Omega\times(0,T]$, in some week
sense~\citep{temam2001} and furnished with appropriate boundary and initial
conditions.
Here, $\bvarepsilon(\bu) := \frac12 (\GRAD \bu + \GRAD \bu ^\top )$ is the
rate-of-strain tensor with $\nu \ge 0$ being the kinematic viscosity of the
fluid.

We are interested in the case where the kinematic viscosity vanishes, \ie
$\nu \rightarrow 0$, which corresponds to the inviscid flow regime. It is
usually believed \citep{Constantin_2015, Kukavica_2022} that the inviscid
limit of the Navier–Stokes equations is given by the Euler equations,
\begin{equation}
  \label{eq:euler}
  \begin{cases}
    \begin{aligned}
      \p_t \bu + \bu \SCAL \GRAD \bu + \GRAD p &= 0, \\
      \DIV \bu &= 0,
    \end{aligned}
  \end{cases}
\end{equation}
at least away from boundaries. However, establishing this limit for bounded
domains with, for example, Dirichlet boundary conditions remains a
challenging problem; see, for instance, \citep{Masmoudi_2007,Kukavica_2022} and
references therein.

For the Navier–Stokes equations, the natural boundary condition on solid
walls is the \emph{no-slip} condition, $\bu = 0$, which enforces both zero
normal and tangential velocity at the boundary. At high Reynolds numbers
this creates a sharp boundary layer \citep{Schlichting_2017}.
In contrast, the natural boundary condition for the Euler equations on
solid walls is the no-penetration condition, $\bu \SCAL \bn = 0$, also
known as the \emph{slip} boundary condition, which prevents flow through
the boundary while allowing tangential motion. Here $\bn$ is the unit
normal vector to the boundary. Consequently, no boundary
layer emerges. This discrepancy between the viscous and inviscid solutions
at the boundary is difficult to reconcile and the reason why the analytical
investigation of the inviscid limit of Navier–Stokes solutions remains
challenging~\citep{Kukavica_2022}.

The so-called Navier-slip boundary condition \citep{Navier_1823} combines
both slip and no-slip conditions:
\begin{equation}
  \label{eq:navier-slip}
  \begin{cases}
    \begin{aligned}
      \bu\SCAL\bn & = 0,
      \\
      \beta \bu \SCAL \bt_k + 2 \nu (\bvarepsilon(\bu) \SCAL \bn) \SCAL \bt_k
      &= 0, \quad k = 1,2,
    \end{aligned}
  \end{cases}
\end{equation}
where $\bt_k$ are orthogonal unit tangent vectors to $\bn$ on the boundary, and
$\beta$ is a friction coefficient. Note that $\beta \rightarrow 0$ recovers
the slip boundary condition, while $\beta \rightarrow \infty$ corresponds
to enforcing a no-slip boundary condition. \cite{Iftimie_2006} proved that
the inviscid limit of the Navier–Stokes equations with slip boundary
conditions indeed converges to solutions of the Euler equations. This is a
significant result with practical implications as slip conditions eliminate
the formation of boundary layers, which is beneficial for numerical
simulations.

Navier-slip boundary conditions \eqref{eq:navier-slip} with various choices
of friction coefficient $\beta$ is used for \emph{wall-modeled} large-eddy
simulation for channel flows, see for instance \citep{Min_Kim_2004,
Bose_2014}. However, it is not clear whether using Navier-slip boundary
conditions is advantageous for bluff body problems, since avoiding boundary
layers may lead to a \emph{potential} flow, \ie an inviscid,
incompressible, steady, and irrotational flow, which yields zero net drag
force on the body \citep{d'Alambert_1752}. This phenomenon is famously
known as d'Alembert's zero-drag paradox.

\subsection{Resolution of d'Alembert's zero-drag paradox}
d'Alembert's paradox remained unresolved for almost 150 years.
\cite{Prandtl_1904} famously proposed a resolution to the paradox, arguing
that viscous effects near solid boundaries cannot be neglected and that the
no-slip boundary condition must be imposed. In other words, this resolution
of the paradox asserts that inviscid models, such as the Euler equations,
cannot be used to accurately calculate the drag or lift forces on a
body~\citep{Prandtl_1904}.

Recently, \cite{Hoffman_2010} proposed a different point of view for
resolving d'Alembert's paradox by arguing that potential flow is unstable
at the separation points in three space dimensions. This instability then
naturally leads to the development of a turbulent flow with substantial
drag caused by streaks of low-pressure streamwise vorticity generated by
the breakdown of the potential flow near the rear separation region
\citep{Hoffman_2010}. This new perspective has opened a novel direction for
turbulent flow simulation, as it eliminates the need to resolve the
computationally expensive Prandtl boundary layer: When numerically
approximating the flow past a body using the inviscid Euler equations with
slip boundary conditions, while initializing the solution as potential
flow, the flow should become unstable. Computational results reported in
\citep{Hoffman_2011, Jansson_2011, Hoffman_2015, Hoffman_2016,
deAbreu_2016} demonstrated that simulating flow past various
objects—including a full car, an airplane, and landing gear—using slip
boundary conditions produced turbulent solutions with drag and lift
coefficients in close agreement with experimental data.

An interesting observation is that these two resolutions appear to
fundamentally contradict each other as they assume an opposite viewpoint on
the role of the boundary layer. Prandtl’s resolution on the one hand is
supported by extensive experimental evidence of flows studied in the
laboratory \citep{Prandtl_1904}, whereas the viewpoint assumed by
\cite{Hoffman_2010} is corroborated by computational
evidence~\citep{Hoffman_2011, Jansson_2011, Hoffman_2015, Hoffman_2016,
deAbreu_2016}. This raises a fundamental question: which of these two
viewpoints should be regarded as the correct resolution of d'Alembert’s
paradox?

\subsection{Surface roughness as a hidden tuning parameter}

To the best of our knowledge, all of the numerical evidence
\citep{Hoffman_2010, Hoffman_2011, Jansson_2011, Hoffman_2015,
Hoffman_2016, deAbreu_2016} in support of \cite{Hoffman_2010} has been
obtained from simulations using the \texttt{Unicorn} solver
\citep{Hoffman_et_al_2012, Hoffman_et_al_2013}, which is built upon the
\texttt{FEniCS} project \citep{Logg_at_al_2012}. \texttt{Unicorn} uses an
inconsistent Galerkin least-squares (GLS) stabilized finite element
discretization on unstructured meshes \citep{Brooks_1982, Johnson_1986}. We
note that a GLS stabilized finite element approximation---as is the case
for any other discretization technique---contains a number of hidden, or
implicit, modeling parameters, which can significantly affect the
computational results. For simulating flow past a cylinder, which is the
subject of our paper, we note the following hidden tuning parameters:
\begin{enumerate}
  \item[(i)]
    Nonuniform artificial viscosity: the GLS stabilization introduces an
    artificial viscosity that depends on the local element size. For
    unstructured meshes, this leads to a nonuniform dissipation,
    potentially affecting the instability of the potential solution.
  \item[(ii)]
    Geometry approximation error: the curvature of the cylinder must be
    resolved accurately. This requires high-order isoparametric mappings to
    reduce the geometric error.
  \item[(iii)]
    Choice of finite element spaces: when using piecewise linear elements,
    the curved boundary of the cylinder is represented as a polygonal
    approximation, introducing additional errors. To reduce both geometry
    and discretization errors, higher-order polynomial spaces are needed.
\end{enumerate}
At the time when \texttt{Unicorn} was conceptualized and implemented,
\texttt{FEniCS} supported only unstructured tetrahedral meshes with
piecewise linear geometric approximations. As a result, none of these
points have been addressed.

It is the purpose of the present paper to investigate numerically whether
and how these hidden tuning parameters can affect the stability of flow
past a cylinder with slip boundary conditions.

\subsection{Our contribution}
The objective of this paper is to reproduce the numerical experiments
originally reported by \cite{Hoffman_2010} for a three dimensional cylinder
geometry while addressing the three items outlined above. For the sake of
comparison we choose to use a similar discretization technique that was
outlined in \citep{Hoffman_2010}, namely a Galerkin least-squares (GLS)
stabilized finite element discretization on unstructured meshes
\citep{Brooks_1982, Johnson_1986}. We want to emphasize that it is not our
objective to develop, or improve on the GLS discretization and
stabilization approach, or judge the merits of it.

In \citep[Sec.~5.2]{Nazarov_2012}, subsonic flow past a circular cylinder
in 2D was investigated using an Euler solver with a slip boundary
condition. Even at extreme mesh resolutions (down to element sizes of
$\approx 10^{-7}$), a significant deviation from potential flow was
observed, indicating the need for more accurate geometric representation
and numerical resolution. We thus base the solver for the current
investigation on the open source finite element toolkit
\texttt{deal.II}~\citep{dealIIcanonical, dealII96}, that supports
quadrilateral and hexahedral meshes along with high-order isoparametric
mappings for curved geometries. Our solver addresses the three key
limitations outlined above: it enables perfectly symmetric, structured
grids, avoiding artificial dissipation from mesh nonuniformity; it supports
high-order manifold descriptions, significantly reducing geometric
approximation errors; it allows for arbitrary high-order finite element
spaces, improving accuracy in both velocity and pressure approximations. We
have made our solver publicly available for verification and further
experimentation \citep{maier_2025_15455164}.

Our main observations that are discussed in detail in
Sections~\ref{sec:numerics} and~\ref{sec:discussion} can be summarized as
follows:
\begin{enumerate}
  \item[(i)]
    GLS simulations of incompressible, high Reynolds number flow past a 2D
    cylinder using slip boundary conditions---while minimizing the geometry
    error (by using a quasi-uniform mesh, high order boundary
    approximation, and isoparametric discretization)---results in a stable
    potential flow profile with vanishing drag and lift; see
    Section~\ref{subse:parameter_study_2d}.
  \item[(ii)]
    The flow profile and resulting drag and lift values can be manipulated
    dramatically by introducing surface roughness; see
    Section~\ref{subse:parameter_study_2d}. We observe quantitative changes
    for flow statistics of 3 to 5 orders of magnitude.
  \item[(iii)]
    We confirm that this behavior carries over into 3D: When minimizing the
    geometry error (by using quasi-uniform mesh, high order boundary
    approximation, and isoparametric discretization) we observe that GLS
    simulations of incompressible, high Reynolds number flow past a 3D
    cylinder using slip boundary conditions results again in a stable
    potential flow profile with vanishing drag and lift; see
    Section~\ref{subse:parameter_study_3d}.
  \item[(iv)]
    We document quantitatively how the stability and resulting drag and
    lift values of 3D potential flow is affected again by the introduction
    of (a) surface roughness (Section~\ref{subse:parameter_study_3d}) and
    (b) mesh distortion
    (Section~\ref{subse:parameter_study_3d_mesh_distortion}).
\end{enumerate}
This indiccates that numerical surface roughness and mesh distortion can be
used as control parameters to manipulate drag and lift forces resulting in
numerical values spanning various orders of magnitude. This casts some
doubt on the predictive capability of the slip boundary condition for wall
modeling in turbulent simulations of incompressible flow.

\subsection{Paper organization}
The remainder of the paper is organized as follows.
In Section~\ref{sec:fem} we summarize the full and inconsistent
Galerkin least-squares (GLS) stabilized finite element discretizations,
our strategy for boundary approximations, and aspects of our solver
implementation. We conduct a series of validation tests, as well as the
aforementioned parameter studies in Section~\ref{sec:numerics}. We
summarize our findings, discuss implications and conclusions in
Section~\ref{sec:discussion}.


\section{Finite element approximation}
\label{sec:fem}
For our numerical parameter studies reported in Section~\ref{sec:numerics}
we use the same discretization technique that was outlined in
\citep{Hoffman_2010}, a Galerkin least-squares (GLS) stabilized
finite element discretization on unstructured meshes. For the sake of
completeness we now summarize the discretization technique. We have made
our implementation publicly available for verification and further
experimentation \citep{maier_2025_15455164}.

We denote by $\calT_h$ a subdivision of $\Omega$ into a finite number of
disjoint elements $K$ such that $\overline\Omega=\cup_{k\in \calT_h}
\overline K$, where $\overline \Omega$ and $\overline K$ denotes the
closure of $\Omega$ and $K$, respectively. Let us consider a family of
shape-regular and conforming meshes $\{\calT_h\}_{h>0}$, where $h$ denotes
the smallest diameter of all triangles of $\calT_h$. Next, we denote by
$\bg_K: \widehat K \mapsto K$ the diffeomorphism that maps the reference
element $\widehat K$ to the real cell $K$. With each mesh $\calT_h$ we
associate the following continuous approximation space:
\begin{equation}
  \label{eq:Xh}
  \begin{aligned}
    \calX_h
    :=&\
    \{v_h: v_h \in \calC^0(\overline \Omega); \, \forall K\in \calT_h, \,
    v_h|_K \circ \bg_K \in \polQ_k(\widehat K) \},
  \end{aligned}
\end{equation}
where $\polQ_k$ is the set of multivariate polynomials of total degree at
most $k\ge1$ defined over $\widehat K$. For every element $K$, we define a
mesh-size $h_K$ as the minimum distance between any pair of its vertices.

\subsection{Space and time discretization of the Navier-Stokes equations}
For high Reynolds numbers, the Navier-Stokes equations \eqref{eq:ns} are
generally convection dominiated. Consequently, finite element
discretizations have to be stabilized in some form. We follow the approach
in \citep{Hoffman_2010} and stabilized the system using the so-called
General Galerkin method. Next, we split the time interval $[0, T]$ into $N$
intervals of variable size: $0=t_0 < t_1 < \ldots < t_N = T$, and let
$\tau^n = t_{n+1}-t_n$ is the local time step. Given previous
approximations $(\bu_h^k, p_h^k) \in \bcalV_h \times \calX_h$ for time
$t_k$, $k<n$, and where $\bcalV_h = [\calX_h]^d$, we construct a finite
element approximation $(\bu_h^n, p_h^n)\in\bcalV_h\times\calX_h$ for time
$t_n$ by solving
\begin{equation}
  \label{eq:g2}
  F(\bu_h^n, p_h^n ;\bv, q) + S(\bu_h^n, p_h^n ;\bv, q) =
  0,\quad\forall(\bv,q)\in\bcalV_h\times\calX_h.
\end{equation}
Here, $F(\bu_h^n, p_h^n)$ is a fully discrete, weak form corresponding to
\eqref{eq:ns}:
\begin{multline}
  \label{eq:gal}
  F(\bu_h^n, p_h^n ;\bv, q) := \big( D_{\tau} \bu_h^n , \bv \big) + \big(
  \bu_h^n \SCAL \GRAD \bu_h^n,\bv \big)- \big( p_h^n,\DIV \bv \big)
  + \big(2\nu \bvarepsilon(\bu_h^n),\bvarepsilon(\bv) \big)
  \\
  + \big( \DIV \bu_h^n,q)
  - (2\nu \bn_h\SCAL\bvarepsilon(\bu_h^n), \bv )_{\partial\Omega} + ( p_h^n,
  \bn_h \SCAL \bv)_{\partial\Omega},
\end{multline}
where $D_{\tau} \bu_h^n$ is the usual backward differentiation formula of
order two (BDF2) to approximate the time derivative of $\bu_h$:
\begin{align*}
  D_{\tau} \bu_h^n := \frac{2\tau^{n} + \tau^{n-1}}{\tau^{n}(\tau^{n} +
  \tau^{n-1})} \bu_h^n
  - \frac{\tau^{n} + \tau^{n-1}}{\tau^{n} \cdot \tau^{n-1}} \bu^{n-1}_h
  + \frac{\tau^{n}}{\tau^{n-1} \cdot (\tau^{n} + \tau^{n-1})} \bu^{n-2}_h.
\end{align*}
Here, we compute the local time step using the following CFL-type condition: 
$
\tau_n := \text{cfl } h_{\min, \Omega} / |\bu_h^n|,
$
where the constant $\text{cfl}=1$, unless otherwise stated, is used
throughout all numerical simulations presented in this paper.

For convection dominated problems, standard Galerkin methods may become
unstable and require stabilization. To address this, various mesh-dependent
stabilization techniques such as the well-known \emph{Galerkin least
squares} (GLS) stabilization~\citep{Brooks_1982, Johnson_1986} have been
introduced. Such stabilization approaches augment the partial differential
equations with mesh-dependent stabilization terms, that help to stabilize
the numerical solution. Following~\citep{Brooks_1982,
Johnson_1986,Hoffman_2010}, we introduce a GLS stabilization term
$S_{\text{GLS}}(\bu_h^n, p_h^n; \bv, q)$:
\begin{multline}
  \label{eq:stab_consistent}
  S_{\text{GLS}}(\bu_h^n, p_h^n  ; \bv, q) :=
  \\
  \sum_{K\in \calT_h}\Big(\delta_{1}^K \big( D_{\tau}\bu_h^n+ \bu_h^n \SCAL
  \GRAD \bu_h^n
  +\nabla p_h^n \big),\, \bu_h^n \SCAL \GRAD \bv + \GRAD q \Big)_K
  \\
   \quad + \sum_{K\in \calT_h} \big( \delta_2^{K} \DIV \bu_h^n, \, \DIV
  \bv \big)_K,
\end{multline}
where the stabilization coefficients are given by
\begin{equation*}
  \delta_1^K = c_1 \Bigg( \frac{1}{\Delta t_n^{2}} +
  \frac{\max_K\,|\bu^{n-1}_h|^2}{h_K^{2}} \Bigg)^{-\frac12}, \quad
  \delta_2^K = c_2 h_K,
\end{equation*}
when $\nu < |\bu^{n-1}_h| h_K$ and
\begin{equation*}
  \delta_1^K = c_1 h_K^2, \quad \delta_2^K = c_2 h_K^2,
\end{equation*}
if $\nu \ge |\bu^{n-1}_h| h_K$. We have set the stabilization parameters to
$c_1=1$, or $2$, and $c_2=1$ in our numerical simulations.

Note that the stabilization term \eqref{eq:stab_consistent} can be
simplified by dropping the terms for the time derivative
leading to a simplified stabilization term~\citep{Hoffman_2007}:
\begin{multline}
  \label{eq:stab_inconsistent}
  S_{\text{G2}}(\bu_h^n, p_h^n ; \bv, q) :=
  \sum_{K\in \calT_h}\Big(\delta_{1}^K \big(
  \bu_h^n \SCAL \GRAD \bu_h^n + \nabla p_h^n \big),\, \bu_h^n \SCAL \GRAD
  \bv + \GRAD q \Big)_K
  \\
   \quad + \sum_{K\in \calT_h} \big( \delta_2^{K} \DIV \bu_h^n, \, \DIV
  \bv \big)_K.
\end{multline}
This form of inconsistent stabilization--- the stabilization term without
time derivatives and without viscous terms--- has been criticized in the
literature, primarily due to its lack of full consistency with the
underlying differential equations. However, it offers a significant
simplification in implementation, particularly by avoiding time-dependent
test spaces, which are often challenging to handle in practice. Despite its
inconsistency, this approach has shown promising performance in turbulent
flow simulations, as demonstrated in \citep{Hoffman_2007}.

We have observed largely similar results for flow past a cylinder
when using $S_{\text{GLS}}$ and
$S_{\text{G2}}$. We have thus opted to use the consistent stabilization
term $S_{\text{GLS}}$ for the majority of numerical experiments. We comment
on some selected computational results using $S_{\text{G2}}$ as well.

\subsection{Boundary conditions and boundary approximation}
\label{sec:numerics_boundary}
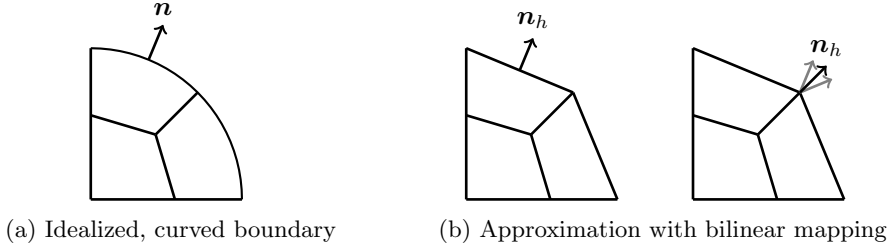
\begin{figure}[t]
  \centering
  \begin{subfigure}[b]{0.4\textwidth}
      \centering
      \begin{tikzpicture}

\draw [black,thick,domain=0:90] plot ({0.0+2.0*cos(\x)}, {0.0+2.0*sin(\x)});

\draw [line width=1] (0,0) -- (2,0);
\draw [line width=1] (0,0) -- (0,2);
\draw [line width=1] (1.11294,0) -- (0.85766,0.85766);
\draw [line width=1] (0, 1.11294) -- (0.85766,0.85766);
\draw [line width=1] (1.41421356237, 1.41421356237) -- (0.85766,0.85766);

\draw [line width=1, ->] (0.7654, 1.8478) -- (0.9567,2.3097) node[above] {$\bn$};

\end{tikzpicture}
      \caption{Idealized, curved boundary}
  \end{subfigure}\hfill
  \begin{subfigure}[b]{0.6\textwidth}
      \centering
      \begin{tikzpicture}
\draw [line width=1] (6+0,0) -- (6+2,0);
\draw [line width=1] (6+0,0) -- (6+0,2);
\draw [line width=1] (6+1.11294,0) -- (6+0.85766,0.85766);
\draw [line width=1] (6+0, 1.11294) -- (6+0.85766,0.85766);
\draw [line width=1] (6+1.41421356237, 1.41421356237) -- (6+0.85766,0.85766);
\draw [line width=1] (6+1.41421356237, 1.41421356237) -- (6+2,0);
\draw [line width=1] (6+1.41421356237, 1.41421356237) -- (6+0,2);

\draw [line width=1, ->] (6+0.7071, 1.7071) -- (6+0.8839,2.1339) node[above] {$\bn_h$};

\draw [line width=1] (9+0,0) -- (9+2,0);
\draw [line width=1] (9+0,0) -- (9+0,2);
\draw [line width=1] (9+1.11294,0) -- (9+0.85766,0.85766);
\draw [line width=1] (9+0, 1.11294) -- (9+0.85766,0.85766);
\draw [line width=1] (9+1.41421356237, 1.41421356237) -- (9+0.85766,0.85766);
\draw [line width=1] (9+1.41421356237, 1.41421356237) -- (9+2,0);
\draw [line width=1] (9+1.41421356237, 1.41421356237) -- (9+0,2);

\draw [line width=1, ->, gray] (9+1.41421356237, 1.41421356237) -- (9+1.41421356237 + 0.1768, 1.41421356237 + 0.4268);

\draw [line width=1, ->, gray] (9+1.41421356237, 1.41421356237) -- (9+1.41421356237 + 0.4268, 1.41421356237 + 0.1768);

\draw [line width=1, ->] (9+1.41421356237, 1.41421356237) -- (9+1.41421356237 + 0.3536, 1.41421356237 + 0.3536) node[above] {$\bn_h$};

\end{tikzpicture}
      \caption{Approximation with bilinear mapping}
  \end{subfigure}
  \caption{(a) Idealized, curved boundary with outward facing boundary
  normal $\bn(\bx)$; (b) corresponding approximation $\bn_h(\bx)$ for the
  case of a boundary approximated with a  bilinear mapping. Here, the
  normal field on each face differs. In order to reconstruct a normal $\bn$
  in the vertex we average the normals coming from all adjacent faces.}
  \label{fig:normal}
\end{figure}
In the numerical experiments outlined below the discretized system
\eqref{eq:g2} is closed by enforcing various boundary conditions. We use
\begin{itemize}
  \item[--] strongly enforced Dirichlet conditions at the inflow
    prescribing the velocity field as
    \begin{align*}
      \bu^n_h(\bx) = \bu_{\text{D}}(\bx),\quad\text{for }\bx
      \in\Gamma_{\text{inlet}},
    \end{align*}
    where $\bu_{\text{D}}(\bx)$ denotes the imposed inflow profile;
  \item[--] weekly enforced Dirichlet conditions on the
    outflow controlling the velocity field via a Nitsche method described
    below,
  \item[--]
    and on the cylinder surface and channel walls, we impose either
    no-slip, $\bu_h^n(\bx)=0$, or slip boundary conditions,
    $\bu_h^n(\bx)\cdot\bn_h = 0$.
\end{itemize}
Here, $\bn_h$ denotes a discrete reconstruction of the outward facing unit
normal field on the boundary $\Gamma=\partial\Omega$.

On the outlet $\Gamma_{\text{outlet}}$ we enforce weakly---via Nitsche's
method---a velocity field to prevent recirculation at high Reynolds
numbers. This is done by augmenting the stabilized equation \eqref{eq:g2}
with a boundary term:
\begin{multline}
  \label{eq:nitsche}
  N(\bu_h^n, p_h^n ;\bv, q) =
  \\
  - \nu(\partial_n \bu_h^n, \bv )_{\Gamma_{\text{outlet}}}
  - \nu(\bu_h^n-\bu_\text{D}^n, \partial_n \bv )_{\Gamma_{\text{outlet}}}
  + \left( \frac{\beta_{\text{D}}}{h} (\bu_h^n-\bu_\text{D}^n),
  \bv\right)_{\Gamma_{\text{outlet}}},
\end{multline}
with $\beta_{\text{D}} = 1$.
Whereas the Dirichlet, slip and no slip boundary conditions are enforced
strongly, meaning they are directly enforced in the finite element function
space rather than through weak imposition or penalty methods. For the slip
boundary condition, we accomplish this by first reconstructing the normal
vector at each collocation point  on the boundary, as described in the
next section. Then, one component of the velocity is expressed as a linear
combination of the others. For example, in 2D, the condition $\bu^n_h \cdot
\bn_h = 0$ expands to $u_{h,0}^n n_0 + u_{h,1}^n n_1 = 0$, which implies
$u_{h,0}^n = -\frac{n_1}{n_0}u_{h,1}^n$. This is done in \texttt{deal.II}
by constructing an additional constrained matrix similar to the process
outlined in \citep[Sec.~3.4]{heltai2021propagating}.

\paragraph{Reconstruction of normal vectors}
For imposing the boundary conditions outlined above we need to reconstruct
an outward facing unit normal field $\bn_h(\bx)$ approximating the normal
field $\bn$ on the boundary $\Gamma=\partial\Omega$. In our numerical
experiments we will use bilinear (or trilinear), as well as higher-order
isoparametric mappings that approximate the (true) curvature of the curved
boundary $\Gamma$ to various polynomial degrees; see
Figure~\ref{fig:normal}. On the interior of a face the discrete normal
$\bn_h(\bx)$ is well defined and we use the exact discrete normal imposed
by the mapping. On collocation points falling on face boundaries (vertices,
or edges), however, the discrete normal field is multi valued. In order to
reconstruct a normal $\bn_h$ in such points we average the various normals
coming from all adjacent faces; see Figure~\ref{fig:normal}(b).

\begin{figure}
  \centering
  \begin{tikzpicture}

\draw [black,thick,domain=225:315] plot ({0.0+0.75*cos(\x)}, {0.0+0.75*sin(\x)});

\draw [line width=1] (-1.5,-1.5) -- (1.5,-1.5);

\draw [line width=1] (-1.5,-1.5) -- (-0.5303,-0.5303);
\draw [line width=1] (+1.5,-1.5) -- (+0.5303,-0.5303);

\filldraw[black] (-1.5,-1.5) circle (2pt);
\filldraw[black] (+1.5,-1.5) circle (2pt);
\filldraw[black] (-0.5303,-0.5303) circle (2pt);
\filldraw[black] (+0.5303,-0.5303) circle (2pt);

\draw [line width=1, ->] (0.0, -0.1) node[] {{\smash{\textbf{cylindrical manifold}}}} -- (0.0,-0.75);

\draw [black,thick,domain=225:315] plot ({3.5+0.75*cos(\x)}, {0.0+0.75*sin(\x)});

\draw [line width=1] (3.5+-1.5,-1.5) -- (3.5+1.5,-1.5);

\draw [line width=1] (3.5-1.5,-1.5) -- (3.5-0.5303,-0.5303);
\draw [line width=1] (3.5+1.5,-1.5) -- (3.5+0.5303,-0.5303);
\draw [line width=1] (3.5+0.0,-1.5) -- (3.5+0.0,-0.75);

\draw [line width=1] (3.5-1.0151,-1.0151) -- (3.5+0.0,-1.125);
\draw [line width=1] (3.5+1.0151,-1.0151) -- (3.5+0.0,-1.125);

\filldraw[gray] (3.5-1.5,-1.5) circle (2pt);
\filldraw[gray] (3.5+1.5,-1.5) circle (2pt);
\filldraw[gray] (3.5-0.5303,-0.5303) circle (2pt);
\filldraw[gray] (3.5+0.5303,-0.5303) circle (2pt);

\filldraw[black] (3.5+0.0,-0.75) circle (2pt);
\filldraw[black] (3.5+0.0,-1.125) circle (2pt);
\filldraw[black] (3.5+0.0,-1.5) circle (2pt);

\filldraw[black] (3.5-1.0151,-1.0151) circle (2pt);
\filldraw[black] (3.5+1.0151,-1.0151) circle (2pt);

\draw [black,thick,domain=225:315] plot ({7+0.75*cos(\x)}, {0.0+0.75*sin(\x)});

\draw [line width=1] (7+-1.5,-1.5) -- (7+1.5,-1.5);

\draw [line width=1] (7-1.5,-1.5) -- (7-0.5303,-0.5303);
\draw [line width=1] (7+1.5,-1.5) -- (7+0.5303,-0.5303);

\filldraw[gray] (7-1.5,-1.5) circle (2pt);
\filldraw[gray] (7+1.5,-1.5) circle (2pt);
\filldraw[gray] (7-0.5303,-0.5303) circle (2pt);
\filldraw[gray] (7+0.5303,-0.5303) circle (2pt);

\filldraw[black] (7-0.6708,-1.5000) circle (2pt);
\filldraw[black] (7+0.6708,-1.5000) circle (2pt);

\filldraw[black] (7-1.2320,-1.2320) circle (2pt);
\filldraw[black] (7-0.7983,-0.7983) circle (2pt);

\filldraw[black] (7+1.2320,-1.2320) circle (2pt);
\filldraw[black] (7+0.7983,-0.7983) circle (2pt);

\filldraw[black] (7-0.2580,-0.7042) circle (2pt);
\filldraw[black] (7+0.2580,-0.7042) circle (2pt);

\filldraw[black] (7-0.5567,-1.2800) circle (2pt);
\filldraw[black] (7-0.3721,-0.9242) circle (2pt);
\filldraw[black] (7+0.5567,-1.2800) circle (2pt);
\filldraw[black] (7+0.3721,-0.9242) circle (2pt);

\draw (3.5, -2.0) node {a) mesh refinement};

\draw (7.0, -2.0) node {b) support points};

\end{tikzpicture}
  \caption{Usage of a high-order manifold and mapping in \texttt{deal.II}:
    The curvature information is used for (a) the generation of new mesh
    points during mesh refinement, and (b) for the placement of
    support points. Here, exemplified for a cubic finite element with
    corresponding cubic bilinear mapping.}
    \label{fig:highordermapping}
\end{figure}
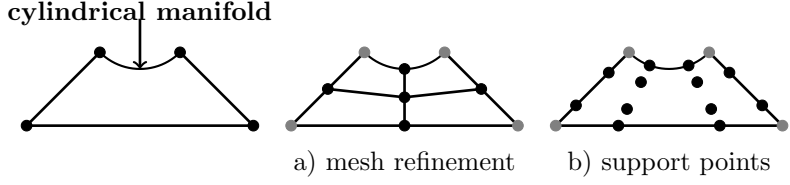
\paragraph{Low-order bilinear (trilinear) and high-order isoparametric
mappings}
\texttt{deal.II} supports various \emph{mapping} strategies for
constructing the approximation space $\calX_h$ via \eqref{eq:Xh}. In
addition a \emph{manifold} object can be attached to the mesh that
describes the curvature of the geometry that is used for mesh refinement,
as well as for constructing the mapping and placement of support points;
see Figure~\ref{fig:highordermapping}. We refer the reader
to~\citep{heltai2021propagating} for a detailed description of the manifold
concept and its implementation in \texttt{deal.II}. Attaching such
high-order manifold allows us to position exactly the new points generated
during mesh refinement and also needed for high-order mapping, as indicated
in Figure~\ref{fig:highordermapping}.

For the sake of completeness we describe briefly how the normal is computed
for a given face. Recall that $\bg_K: \widehat K \mapsto K$ denotes the
mapping between the reference element $\widehat K$ and the real cell $K$.
Let $(\polJ_{K}(\widehat{\bx}))_{ij}= \partial_{\widehat{x}_j}(\bg_K)_i$
denote the Jacobian of $\bg_K(\widehat{x})$. The normal $\bn_h(\bx)$ on a
face is now given by
\begin{align*}
  \bn_h(\bx) = \frac{\polJ_{K}^{-T}(\widehat{\bx})\widehat{\bn}(\widehat{\bx})}
  {\big\|\polJ_{K}^{-T}(\widehat{\bx})\widehat{\bn}(\widehat{\bx})\big\|},
\end{align*}
where $\widehat{\bx} = \bg_K^{-1}(\bx)$, and $\widehat{\bn}$ is the
corresponding normal of $\partial\widehat{K}$ in $\widehat{\bx}$.

\subsection{Solution procedure and linear solver}
Given previous
approximations $(\bu_h^k, p_h^k) \in \bcalV_h \times \calX_h$ for time
$t_k$, $k<n$, we need to find the solution $(\bu_h^n, p_h^n)$ to the
nonlinear equation \eqref{eq:g2}, viz.,
$\forall(\bv,q)\in\bcalV_h\times\calX_h:$
\begin{equation*}
  B(\bu_h^n, p_h^n; \bv, q) := F(\bu_h^n, p_h^n; \bv, q) + S(\bu_h^n,
  p_h^n; \bv, q) + N(\bu_h^n, p_h^n ;\bv, q)
  \;=\;
  0.
\end{equation*}
Starting from an initial guess $(\bu_h^{(0)}, p_h^{(0)})$ for $(\bu_h^{n},
p_h^{n})$ we construct a sequence of approximations with a Newton
iteration:
\begin{align}
  \label{eq:newton_step}
  B'_{\bu,p}(\bu^{(k)},p^{(k)};\bv, q)[\delta\bu_h^n,\delta p_h^n]
  &= - B(\bu^{(k)}, p^{(k)};\bv, q),
  \\[0.25em]
  \bu^{(k)} &= \bu^{(k)} + \delta\bu^{(k)},
  \notag
  \\
  p^{(k)} &= p^{(k)} + \delta p^{(k)}.
  \notag
\end{align}
Here, $B'_{\bu,p}(\bu^{\ast}, p^{\ast};\bv, q)[\delta\bu,\delta p]$ denotes
the Gâteaux derivative of the bilinear form $B(.;.)$ at point $(\bu^{\ast},
p^{\ast})$ in direction $(\delta\bu,\delta p)$ with respect to the first
argument. We obtain:
\begin{multline*}
    F'_{\bu,p}(\bu^{\ast}, p^{\ast};\bv, q)[\delta\bu,\delta p] =
    \left(D'_{\tau}\delta\bu + \bu^* \SCAL \GRAD \delta \bu + \delta \bu
    \SCAL \GRAD \bu^*, \bv \right)
    \\
    - \big(\delta p, \DIV \bv \big)
    + \big(2\nu\bvarepsilon(\delta\bu), \bvarepsilon(\bv)\big)
    + \big( \DIV \delta\bu, q),
\end{multline*}
where
\begin{align*}
  D'_{\tau}\bu
  \,=\,
  \frac{2\tau^{n} + \tau^{n-1}}{\tau^{n}(\tau^{n} + \tau^{n-1})}\,\bu.
\end{align*}
Correspondingly, the derivative of the GLS stabilization term reads
\begin{multline*}
  S'_{\bu,p}(\bu^{\ast}, p^{\ast};\bv, q)[\delta\bu,\delta p] =
  \\
  \sum_{K\in \calT_h}\Big(\delta_{1}^K \big(D'_{\tau}\delta\bu +
  \bu^{\ast}\SCAL\GRAD\delta\bu + \delta\bu\SCAL\GRAD\bu^{\ast}
  +\nabla\delta p\big),\, \bu^{\ast} \SCAL \GRAD \bv + \GRAD q \Big)_K
  \\
  +\sum_{K\in \calT_h}\Big(\delta_{1}^K \big(D'_{\tau}\bu^{\ast} +
  \bu^{\ast}\SCAL\GRAD\bu^{\ast}
  +\nabla p^{\ast}\big),\, \delta\bu \SCAL \GRAD \bv \Big)_K
  + \sum_{K\in \calT_h} \big( \delta_2^{K} \DIV \delta\bu, \, \DIV
  \bv \big)_K.
\end{multline*}
The linear subsystem \eqref{eq:newton_step} is solved with an iterative
GMRES method. We precondition the subsystems with a monolithic geometric
multigrid using point-Jacobi iterations as smoother accelerated with a
relaxation scheme. On the coarsest level a direct solver from the
\texttt{Amesos} package from the \texttt{Trilinos} library is
used~\citep{heroux2003overview}). For more informations on the linear
solver, we refer interested readers to~\citep{Prieto2024}. As stopping
criterion, we use an absolute tolerance of $10^{-7}$ for the nonlinear
solver and an relative tolerance of $10^{-2}$ for the linear solver. To
keep the computational efficiency high, we evaluate the linear subsystem
\eqref{eq:newton_step} with a matrix-free
approach~\citep{kronbichler2012generic}. Finally, the entire system is set
up fully MPI parallelized~\citep{dealIIcanonical,dealII96}.


\subsection{Computational cost analysis}
We briefly analyze various computational costs associated with our choice
of discretization. We compare using $\mathbb{Q}_2$ and $\mathbb{Q}_1$
ansatz spaces, as well as using an isoparametric mapping instead of a
linear mapping. As a benchmark, we run the ``flow past a 2D cylinder with
$Re=100$'' (see Section~\ref{subse:validation_turek} configuration with
refinement levels $r=4,5$ and $\text{cfl}=1,2$. We specify the number of
cells, number of degrees of freedom, total simulation time, number of time
steps, average number of non-linear iterations, average number of linear
iterations, and simulation time normalized by the total number of linear
iterations. All simulations have been run with 12 MPI ranks on a
workstation featuring an Intel Core i9-14900.
\begin{table}[t]
  \centering
  \begin{tabular}{lcccc}
    \toprule
    &$\mathbb{Q}_1$ ($r=4$) & $\mathbb{Q}_1$ ($\text{cfl}=2$) &
    $\mathbb{Q}_2$ & $\mathbb{Q}_2$ (lin.)
    \\[0.25em]
    number of cells & 22,528 & 90,112 & 22,528 & 22,528 \\
    number of DoFs & 68,976 & 273,120 & 273,120 & 273,120\\
    tot. simulation time [s] & 340.5 &	1170	 & 960.1	 & 944.5  \\
    number of time steps & 10806 &	10926 &	10831 &	10830 \\
    avg. nonlinear iterations & 2.18	& 2.00 & 	2.94	 & 2.94\\
    avg. linear iterations & 1.63 & 2.00 & 2.05 & 2.10 \\
    norm. sim. time [ms] & 8.87 & 26.70 & 	14.73 & 14.16 \\
    \bottomrule
  \end{tabular}
  \caption{Comparison of runtime of the ``flow past a 2D cylinder with
    $Re=100$'' configuration (Section~\ref{subse:validation_turek}): all
    variants are run with $r=5$, $\text{cfl}=1$, and isoparametric mapping,
    except when otherwise indicated: the second $\mathbb{Q}_1$ computation
    on refinement level $r=5$ is computed with $\text{cfl}=2$ for direct
    comparison with $\mathbb{Q}_2$; the second $\mathbb{Q}_2$ variant is
    computed with linear mapping.}
  \label{table:computational_cost}
\end{table}
The results are summarized in Table~\ref{table:computational_cost}. They
indicate that running $\mathbb{Q}_2$ is about $20\%$ faster that running
$\mathbb{Q}_1$ with the same number of unknowns,, which is mainly related
to the fact that the throughput of matrix-free operator evaluation
increases with the polynomial degree~\citep{kronbichler2012generic}. The
timing differences between isoparametric and linear mappings are minor with
a deviation of less than $1\,\%$). This is due to the implementation of
boundary mappings found in \texttt{deal.II}, where geometric metrics are
precomputed at quadrature-point level so that the same amount of data needs
to be loaded from main memory.


\section{Numerical validation and experiments}
\label{sec:numerics}
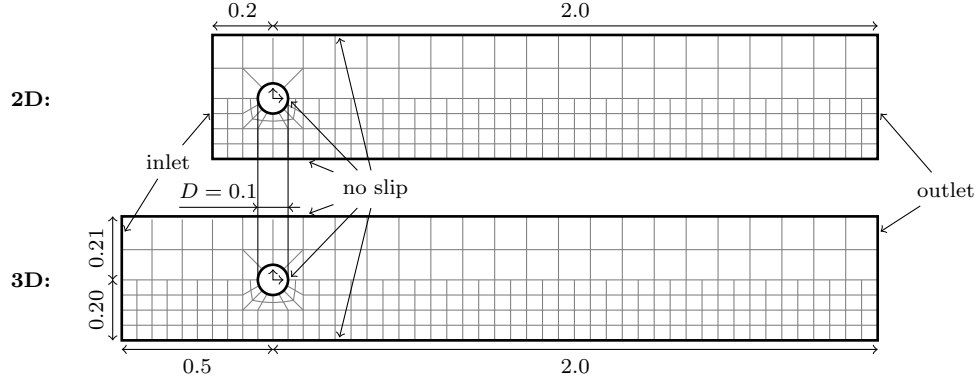
\begin{figure}[!t]
\centering
  \newcommand{\scalefpc}{4.0}
\newcommand{\shiftfpc}{-0.6*\scalefpc}
\begin{tikzpicture}
\draw[gray](-0.1*\scalefpc, -0.2*\scalefpc) -- (-0.1*\scalefpc,0.21*\scalefpc);
\draw[gray](0.1*\scalefpc, -0.2*\scalefpc) -- (0.1*\scalefpc,0.21*\scalefpc);
\draw[gray](-0.2*\scalefpc, +0.1*\scalefpc) -- (2.0*\scalefpc,+0.1*\scalefpc);
\draw[gray](-0.2*\scalefpc, -0.1*\scalefpc) -- (2.0*\scalefpc,-0.1*\scalefpc);
\draw[gray](-0.2*\scalefpc, -0.15*\scalefpc) -- (2.0*\scalefpc,-0.15*\scalefpc);

\draw[gray](-0.2*\scalefpc, 0.0*\scalefpc) -- (-0.05*\scalefpc,0.0*\scalefpc);
\draw[gray](+0.05*\scalefpc, 0.0*\scalefpc) -- (+2.00*\scalefpc,0.0*\scalefpc);

\draw[gray](-0.2*\scalefpc, -0.05*\scalefpc) -- (-0.1*\scalefpc,-0.05*\scalefpc);
\draw[gray](+0.1*\scalefpc, -0.05*\scalefpc) -- (+2.00*\scalefpc,-0.05*\scalefpc);
\draw[gray](-0.0*\scalefpc, 0.05*\scalefpc) -- (-0.0*\scalefpc,0.1*\scalefpc);
\draw[gray](-0.0*\scalefpc, -0.05*\scalefpc) -- (-0.0*\scalefpc,-0.1*\scalefpc);
\draw[gray](+0.03535533905*\scalefpc, +0.03535533905*\scalefpc) -- (+0.1*\scalefpc,+0.1*\scalefpc);
\draw[gray](-0.03535533905*\scalefpc, +0.03535533905*\scalefpc) -- (-0.1*\scalefpc,+0.1*\scalefpc);
\draw[gray](+0.03535533905*\scalefpc, -0.03535533905*\scalefpc) -- (+0.1*\scalefpc,-0.1*\scalefpc);
\draw[gray](-0.03535533905*\scalefpc, -0.03535533905*\scalefpc) -- (-0.1*\scalefpc,-0.1*\scalefpc);

\draw[gray](-0.1*\scalefpc,-0.05*\scalefpc) -- (-0.046194*\scalefpc, -0.019134*\scalefpc);
\draw[gray](-0.05*\scalefpc,-0.1*\scalefpc) -- (-0.019134*\scalefpc, -0.046194*\scalefpc);
\draw[gray](+0.05*\scalefpc,-0.1*\scalefpc) -- (+0.019134*\scalefpc, -0.046194*\scalefpc);
\draw[gray](+0.1*\scalefpc,-0.05*\scalefpc) -- (+0.046194*\scalefpc, -0.019134*\scalefpc);

\draw[gray](-0.075*\scalefpc,-0.0*\scalefpc)
-- (-0.073097*\scalefpc, -0.034567*\scalefpc)
-- (-0.067678*\scalefpc, -0.067678*\scalefpc)
-- (-0.034567*\scalefpc, -0.073097*\scalefpc)
-- (-0.000000*\scalefpc, -0.075000*\scalefpc);

\draw[gray](+0.075*\scalefpc,-0.0*\scalefpc)
-- (+0.073097*\scalefpc, -0.034567*\scalefpc)
-- (+0.067678*\scalefpc, -0.067678*\scalefpc)
-- (+0.034567*\scalefpc, -0.073097*\scalefpc)
-- (+0.000000*\scalefpc, -0.075000*\scalefpc);

\foreach \i in {1,...,17}
{
\draw[gray](+0.1*\scalefpc + 1.9/18.0*\i*\scalefpc, +0.0*\scalefpc) -- (+0.1*\scalefpc + 1.9/18.0*\i*\scalefpc,+0.21*\scalefpc);
}

\foreach \i in {1,...,35}
{
\draw[gray](+0.1*\scalefpc + 1.9/36.0*\i*\scalefpc, +0.0*\scalefpc) -- (+0.1*\scalefpc + 1.9/36.0*\i*\scalefpc,-0.2*\scalefpc);
}

\foreach \i in {1}
{
\draw[gray](-0.2*\scalefpc + 0.1/2.0*\i*\scalefpc, +0.0*\scalefpc) -- (-0.2*\scalefpc + 0.1/2.0*\i*\scalefpc,-0.2*\scalefpc);
}

\foreach \i in {1}
{
\draw[gray](-0.1*\scalefpc + 0.2/2.0*\i*\scalefpc, +0.1*\scalefpc) -- (-0.1*\scalefpc + 0.2/2.0*\i*\scalefpc,+0.2*\scalefpc);
}

\foreach \i in {1,...,3}
{
\draw[gray](-0.1*\scalefpc + 0.2/4.0*\i*\scalefpc, -0.1*\scalefpc) -- (-0.1*\scalefpc + 0.2/4.0*\i*\scalefpc,-0.2*\scalefpc);
}

\draw  [line width=1pt] (0*\scalefpc,0*\scalefpc) circle [blue, radius=0.05*\scalefpc];

\draw[draw=black, line width=1pt] (-0.2*\scalefpc,-0.2*\scalefpc) rectangle ++(2.2*\scalefpc,0.41*\scalefpc);

\draw[gray](-0.1*\scalefpc, -0.2*\scalefpc+ \shiftfpc) -- (-0.1*\scalefpc,0.21*\scalefpc+ \shiftfpc);
\draw[gray](0.1*\scalefpc, -0.2*\scalefpc+ \shiftfpc) -- (0.1*\scalefpc,0.21*\scalefpc+ \shiftfpc);

\draw[gray](-0.5*\scalefpc, +0.1*\scalefpc+ \shiftfpc) -- (2.0*\scalefpc,+0.1*\scalefpc+ \shiftfpc);
\draw[gray](-0.5*\scalefpc, -0.1*\scalefpc+ \shiftfpc) -- (2.0*\scalefpc,-0.1*\scalefpc+ \shiftfpc);
\draw[gray](-0.5*\scalefpc, -0.15*\scalefpc+ \shiftfpc) -- (2.0*\scalefpc,-0.15*\scalefpc+ \shiftfpc);

\draw[gray](-0.5*\scalefpc, 0.0*\scalefpc+ \shiftfpc) -- (-0.05*\scalefpc,0.0*\scalefpc+ \shiftfpc);
\draw[gray](+0.05*\scalefpc, 0.0*\scalefpc+ \shiftfpc) -- (+2.00*\scalefpc,0.0*\scalefpc+ \shiftfpc);

\draw[gray](-0.5*\scalefpc, -0.05*\scalefpc+ \shiftfpc) -- (-0.1*\scalefpc,-0.05*\scalefpc+ \shiftfpc);
\draw[gray](+0.1*\scalefpc, -0.05*\scalefpc+ \shiftfpc) -- (+2.00*\scalefpc,-0.05*\scalefpc+ \shiftfpc);

\draw[gray](-0.0*\scalefpc, 0.05*\scalefpc+ \shiftfpc) -- (-0.0*\scalefpc,0.1*\scalefpc+ \shiftfpc);
\draw[gray](-0.0*\scalefpc, -0.05*\scalefpc+ \shiftfpc) -- (-0.0*\scalefpc,-0.1*\scalefpc+ \shiftfpc);
\draw[gray](+0.03535533905*\scalefpc, +0.03535533905*\scalefpc+ \shiftfpc) -- (+0.1*\scalefpc,+0.1*\scalefpc+ \shiftfpc);
\draw[gray](-0.03535533905*\scalefpc, +0.03535533905*\scalefpc+ \shiftfpc) -- (-0.1*\scalefpc,+0.1*\scalefpc+ \shiftfpc);
\draw[gray](+0.03535533905*\scalefpc, -0.03535533905*\scalefpc+ \shiftfpc) -- (+0.1*\scalefpc,-0.1*\scalefpc+ \shiftfpc);
\draw[gray](-0.03535533905*\scalefpc, -0.03535533905*\scalefpc+ \shiftfpc) -- (-0.1*\scalefpc,-0.1*\scalefpc+ \shiftfpc);

\draw[gray](-0.1*\scalefpc,-0.05*\scalefpc+ \shiftfpc) -- (-0.046194*\scalefpc, -0.019134*\scalefpc+ \shiftfpc);
\draw[gray](-0.05*\scalefpc,-0.1*\scalefpc+ \shiftfpc) -- (-0.019134*\scalefpc, -0.046194*\scalefpc+ \shiftfpc);
\draw[gray](+0.05*\scalefpc,-0.1*\scalefpc+ \shiftfpc) -- (+0.019134*\scalefpc, -0.046194*\scalefpc+ \shiftfpc);
\draw[gray](+0.1*\scalefpc,-0.05*\scalefpc+ \shiftfpc) -- (+0.046194*\scalefpc, -0.019134*\scalefpc+ \shiftfpc);

\draw[gray](-0.075*\scalefpc,-0.0*\scalefpc+ \shiftfpc)
-- (-0.073097*\scalefpc, -0.034567*\scalefpc+ \shiftfpc)
-- (-0.067678*\scalefpc, -0.067678*\scalefpc+ \shiftfpc)
-- (-0.034567*\scalefpc, -0.073097*\scalefpc+ \shiftfpc)
-- (-0.000000*\scalefpc, -0.075000*\scalefpc+ \shiftfpc);

\draw[gray](+0.075*\scalefpc,-0.0*\scalefpc+ \shiftfpc)
-- (+0.073097*\scalefpc, -0.034567*\scalefpc+ \shiftfpc)
-- (+0.067678*\scalefpc, -0.067678*\scalefpc+ \shiftfpc)
-- (+0.034567*\scalefpc, -0.073097*\scalefpc+ \shiftfpc)
-- (+0.000000*\scalefpc, -0.075000*\scalefpc+ \shiftfpc);

\foreach \i in {1,...,17}
{
\draw[gray](+0.1*\scalefpc + 1.9/18.0*\i*\scalefpc, +0.0*\scalefpc+ \shiftfpc) -- (+0.1*\scalefpc + 1.9/18.0*\i*\scalefpc,+0.21*\scalefpc+ \shiftfpc);
}

\foreach \i in {1,...,35}
{
\draw[gray](+0.1*\scalefpc + 1.9/36.0*\i*\scalefpc, +0.0*\scalefpc+ \shiftfpc) -- (+0.1*\scalefpc + 1.9/36.0*\i*\scalefpc,-0.2*\scalefpc+ \shiftfpc);
}

\foreach \i in {1,...,3}
{
\draw[gray](-0.5*\scalefpc + 0.1/1.0*\i*\scalefpc, +0.0*\scalefpc+ \shiftfpc) -- (-0.5*\scalefpc + 0.1/1.0*\i*\scalefpc,+0.2*\scalefpc+ \shiftfpc);
}

\foreach \i in {1,...,7}
{
\draw[gray](-0.5*\scalefpc + 0.1/2.0*\i*\scalefpc, +0.0*\scalefpc+ \shiftfpc) -- (-0.5*\scalefpc + 0.1/2.0*\i*\scalefpc,-0.2*\scalefpc+ \shiftfpc);
}

\foreach \i in {1}
{
\draw[gray](-0.1*\scalefpc + 0.2/2.0*\i*\scalefpc, +0.1*\scalefpc+ \shiftfpc) -- (-0.1*\scalefpc + 0.2/2.0*\i*\scalefpc,+0.2*\scalefpc+ \shiftfpc);
}

\foreach \i in {1,...,3}
{
\draw[gray](-0.1*\scalefpc + 0.2/4.0*\i*\scalefpc, -0.1*\scalefpc+ \shiftfpc) -- (-0.1*\scalefpc + 0.2/4.0*\i*\scalefpc,-0.2*\scalefpc+ \shiftfpc);
}

\draw [line width=1pt] (0*\scalefpc,\shiftfpc) circle [blue, radius=0.05*\scalefpc];

\draw[draw=black,line width=1pt] (-0.5*\scalefpc,-0.2*\scalefpc + \shiftfpc) rectangle ++(2.5*\scalefpc,0.41*\scalefpc);


\draw[black](-0.05*\scalefpc, 0*\scalefpc+ \shiftfpc) -- (-0.05*\scalefpc,0.0);

\draw[black](+0.05*\scalefpc, 0*\scalefpc+ \shiftfpc) -- (+0.05*\scalefpc,0.0);

\draw[black,<->](-0.50*\scalefpc, -0.23*\scalefpc+ \shiftfpc) -- node[below] {\footnotesize 0.5} ++ (+0.50*\scalefpc, 0.0);

\draw[black,<->](+0.00*\scalefpc, -0.23*\scalefpc+ \shiftfpc) -- node[below] {\footnotesize 2.0} ++ (+2.00*\scalefpc, 0.0);

\draw[black,<->](-0.20*\scalefpc, +0.24*\scalefpc) -- node[above] {\footnotesize 0.2} ++ (+0.20*\scalefpc, 0.0);

\draw[black,<->](+0.00*\scalefpc, +0.24*\scalefpc) -- node[above] {\footnotesize 2.0} ++ (+2.00*\scalefpc, 0.0);

\draw[black,<->](-0.53*\scalefpc, +0.0*\scalefpc+ \shiftfpc) -- node[above, rotate=90] {\footnotesize 0.21} ++ (+0.00*\scalefpc, 0.21*\scalefpc);

\draw[black,<->](-0.53*\scalefpc, +0.0*\scalefpc+ \shiftfpc) -- node[above, rotate=90] {\footnotesize 0.20} ++ (+0.00*\scalefpc, -0.20*\scalefpc);

\node at (-0.8*\scalefpc, 0.0) {\textbf{\footnotesize 2D:}};

\node at (-0.8*\scalefpc, \shiftfpc) {\textbf{\footnotesize 3D:}};

\node [anchor=west] (A) at (0.2*\scalefpc, 0.5*\shiftfpc){\footnotesize no slip};

\draw[black, ->](A) -- (+0.06*\scalefpc,-0.01*\scalefpc);
\draw[black, ->](A) -- (+0.22*\scalefpc, +0.2*\scalefpc);
\draw[black, ->](A) -- (+0.12*\scalefpc, -0.21*\scalefpc);

\draw[black, ->](A) -- (+0.06*\scalefpc,+0.01*\scalefpc+ \shiftfpc);
\draw[black, ->](A) -- (+0.22*\scalefpc, -0.19*\scalefpc+ \shiftfpc);
\draw[black, ->](A) -- (+0.12*\scalefpc, +0.22*\scalefpc+ \shiftfpc);

\node [anchor=west] (B) at (-0.45*\scalefpc, 0.35*\shiftfpc){\footnotesize inlet};

\draw[black, ->](B) -- (-0.49*\scalefpc,+0.16*\scalefpc+ \shiftfpc);

\draw[black, ->](B) -- (-0.21*\scalefpc,-0.05*\scalefpc);

\node [anchor=west] (C) at (2.1*\scalefpc, 0.5*\shiftfpc){\footnotesize outlet};

\draw[black, ->](C) -- (2.01*\scalefpc,+0.16*\scalefpc+ \shiftfpc);

\draw[black, ->](C) -- (2.01*\scalefpc,-0.05*\scalefpc);


\draw[black, <->](C) (0.0*\scalefpc, 0.035*\scalefpc) -- (0.0*\scalefpc, 0.0*\scalefpc) -- (0.035*\scalefpc,0.0*\scalefpc);

\draw[black, <->](C) (0.0*\scalefpc, 0.035*\scalefpc+ \shiftfpc) -- (0.0*\scalefpc, 0.0*\scalefpc+ \shiftfpc) -- (0.035*\scalefpc,0.0*\scalefpc+ \shiftfpc);

\draw[black,->](-0.31*\scalefpc, 0.60*\shiftfpc) -- node[above] {\footnotesize $D=0.1$} ++ (+0.26*\scalefpc, 0.0);

\draw[black,<-](0.05*\scalefpc, 0.60*\shiftfpc) -- node[above] {} ++ (+0.04*\scalefpc, 0.0);

\draw[black](-0.05*\scalefpc, 0.60*\shiftfpc) -- node[above] {} ++ (+0.1*\scalefpc, 0.0);

\end{tikzpicture}
  \caption{%
    2D and 3D computational domains and chosen boundary conditions for the
    validation tests. A cylindrical cutout is centered at the origin with a
    diameter of $D=0.1$. In the background, computational meshes are shown:
    the coarse mesh is shown on the top half and a single-refined mesh is
    shown on the bottom half. For 3D, only a 2D $xy$ cut plane is shown;
    the actual 3D mesh is obtained by extruding the 2D mesh in the third
    dimension to a depth of with $L=0.41$ with a total of 4 subdivisions on
    the coarsest level.}
  \label{fig:geometries}
\end{figure}
We now present a small number of verification and validation tests for our
solver that are all based on various established ``flow past a cylinder''
benchmarks with different Reynolds numbers that have been introduced over
the last decades \citep{Schafer_1996}. To this end we consider two
geometries, one for 2D and one for 3D simulations.
Figure~\ref{fig:geometries} shows the geometries together with the coarsest
mesh level and the first refinement level. Unless noted otherwise, we
perform all numerical experiments with one of the two geometries and
(coarse) meshes shown in Figure~\ref{fig:geometries}. Finer meshes are
obtained from the two coarse meshes by globally refining the mesh with an
appropriate curvature approximation; see
Section~\ref{sec:numerics_boundary}.
We choose to use the GLS stabilization term $S_{\text{GLS}}(\bu_h^n, p_h^n
; \bv, q)$ given by \eqref{eq:stab_consistent} throughout.
\paragraph{Quantities of interest}
Let $\Gamma_{\text{cyl}}$ denote the boundary of the cylindrical cutout for
the 2D or 3D domain; see Figure~\ref{fig:geometries}. In analogy to
\citep{Schafer_1996} we introduce drag and lift coefficients as follows:
\begin{align}
  \label{eq:drag_lift}
  c_{\text{d}} = \frac{2}{A \bar{u}^2} F_x,
  \quad
  c_{\text{l}} = \frac{2}{A \bar{u}^2} F_y,
  \quad\text{where }
  \bF = \int_{\Gamma_{\text{cyl}}} \big(- p \bI + 2\nu \bvarepsilon (\bu)
  \big)\cdot \bn \ud S.
\end{align}
Here, $A$ denotes the cross section of the cylinder, which is $A=D=0.1$ in
2D, and $A=DL=0.1\cdot0.41$ in 3D. The quantity $\bar{u}$ denotes the mean
inflow velocity. As a last quantity of interest we introduce the pressure
difference $\Delta p$ between the front and back of the cylinder
\citep{Schafer_1996}.

\subsection{Validation: flow past a 2D cylinder with $Re=20,\;100$}
\label{subse:validation_turek}
We start by computing a well-known benchmark configuration introduced in
\citep{Schafer1996}. The geometry and boundary conditions are the same as
the 2D mesh depicted in Figure~\ref{fig:geometries}. We use no
slip boundary conditions on cylinder and walls with a parabolic inflow
profile,
\begin{align*}
  \bu_{\text{D}}(\bx) = \big(4\,u_{\text{max}}\,x_2 (H-x_2)/H^2, \, 0\big),
\end{align*}
where $H = 0.41$. We set the kinematic viscosity to $\nu=0.001$ and
$u_{\text{max}} = 0.3$, and $u_{\text{max}} = 1.5$, respectively.
This translates to Reynolds numbers $Re = 20$ with $\bar u=0.2$, and
$Re = 100$ with $\bar u = 1.0$, respectively. The simulation is run from
time $t=0$ to $t=20$ with isoparametric, continuous $\polQ_1$ finite
elements; see Section~\ref{sec:numerics_boundary}. We
perform a convergence study on a sequence of increasingly globally refined
grids with refinement levels $r=4$ to $r=7$ amounting to a total of $69k$
to $4.34M$ degrees of freedom. We set the stabilization parameters to
$c_1=2$, and $c_2=1$.
\begin{table}[t]
  \centering
  \begin{tabular}{lcccc}
    \toprule
      & {\bfseries minimum} & {\bfseries maximum}
      & {\bfseries average $\mu$}    & {\bfseries deviation $\sigma$} \\[0.25em]
      $c_d$      & \mynum{3.1522}  & \mynum{3.2155}  & \mynum{3.1840}  & \mynum{0.0220}  \\
      $c_l$      & \mynum{-1.0173} & \mynum{ 0.9841} & \mynum{-0.0187} & \mynum{ 0.7055} \\
      $\Delta p$ & \mynum{2.4201}  & \mynum{2.5069}  & \mynum{2.4637}  & \mynum{0.0299}  \\
    \bottomrule
  \end{tabular}
  \caption{Validation: flow past a 2D cylinder at $Re=100$ with no-slip
    boundary conditions. Minimum, maximum, temporal average and standard
    deviation of drag and lift coefficients, $c_d$, $c_l$, and pressure
    difference $\Delta p$ are computed on the time interval $[5,20]$
    for refinement level $r=7$.}
  \label{table:Re_100}
\end{table}
\begin{figure}[t]
  \centering
  \begin{subfigure}[b]{0.9\textwidth}
      \centering
      \includegraphics[trim=0 0 0 0, clip, width=0.9\textwidth]{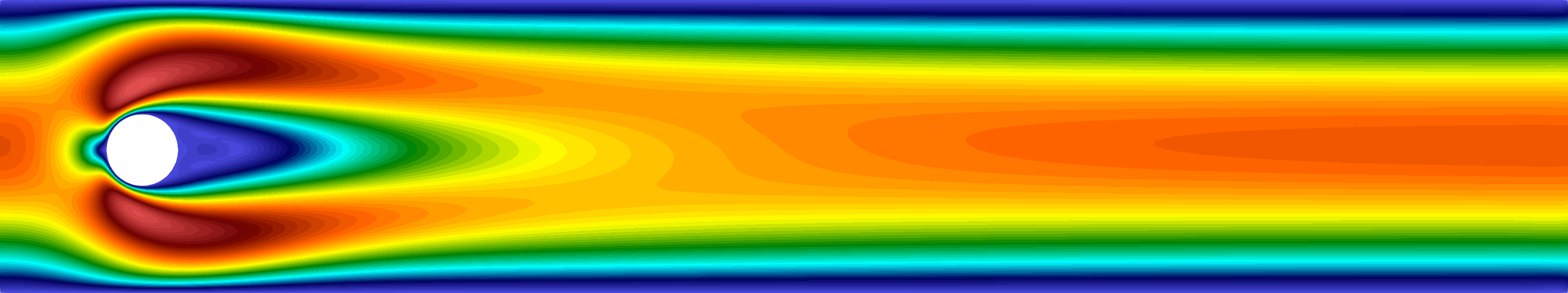}
      \caption{$Re=20$}
  \end{subfigure}\hfill
  \begin{subfigure}[b]{0.9\textwidth}
      \centering
      \includegraphics[trim=0 0 0 0, clip, width=0.9\textwidth]{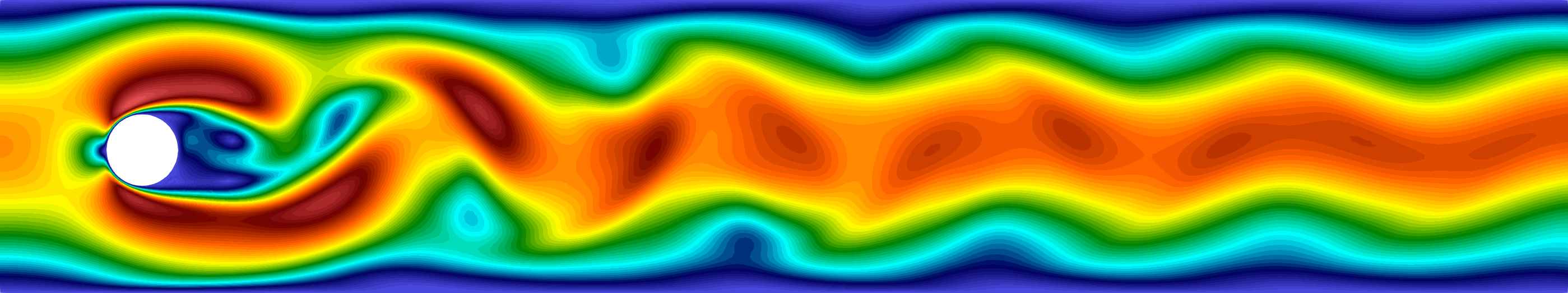}
      \caption{$Re=100$}
  \end{subfigure}\hfill
  \caption{Validation: flow past a 2D cylinder at $Re=20$ and $100$ with
    no-slip boundary conditions. Temporal snapshot of the velocity magnitudes
    at time $T = 20$ are shown for both steady and unsteady simulations
    with $Re=20$, and $Re=100$, respectively. A no-slip boundary condition
    is applied on both the cylinder surface and the channel walls.
    The shown snapshots are for a simulation using
    $\polQ_1$ finite elements on a coarse grid with refinement level $r=4$,
    totaling $69k$ degrees of freedom.}
  \label{cyl:Turek_Re_20_100}
\end{figure}
Temporal snapshots of velocity magnitudes for time $T = 20$ for both
Reynolds numbers are shown in Figure~\ref{cyl:Turek_Re_20_100}.
For the $Re=20$ case we observe a steady flow with
$c_d=5.57819(63)$, $c_l=0.010603(24)$, and $\Delta p=0.1175220(10)$. The
values and uncertainty intervals have been obtained by extrapolating the
values for the four refinement levels to the limit. The results are in good
agreement with \citep[Table~3]{Schafer1996} and \citep{Nabh1998}. The corresponding
results for the unsteady case with $Re = 100$ are summarized in
Table~\ref{table:Re_100} for refinement level $r=7$.
Here, for a quantity of interest $q$ we introduce a temporal average
$\mu_q$ and standard deviation $\sigma_q$:
\begin{align}
  \label{eq:average_deviation}
  \mu_q\;:=\;
  \frac{1}{t_2-t_1}\int_{t_1}^{t_2}q(t)\,\text{d}t,
  \qquad
  \sigma_q^2\;:=\;
  \frac{1}{t_2-t_1}\int_{t_1}^{t_2}\big(q(t) - \mu_q\big)^2\,\text{d}t.
\end{align}
We again observe that our results are in good agreement with the values
reported in \citep[Table~4]{Schafer1996}.

\subsection{Validation: flow past a 3D cylinder with $Re=3900$}
Next we perform a 3D validation test with the 3D geometry outlined in
Figure~\ref{fig:geometries}, again with no slip conditions on the cylinder,
but with slip boundary conditions at the walls of the channel
\citep{hoffman2009}. Furthermore, the cylinder is now inscribed
symmetrically with equal distance to the top and bottom wall. The kinematic
viscosity is set to $\nu=0.001$ and the input profile is chosen to be a
constant inflow at $\bar u = u_{\text{max}}=39$. This translates to a
Reynolds number of $Re = 3900$. The simulation is run from time $t=0$ until
we hit an asymptotic statistical regime at $t=0.5$. Zero initial
data is used for the flow combined with a linear ramp-up of the inflow from $0$ to $u_{\text{max}}$
for $t=0$ to $t=0.02$. We choose an isoparametric, continuous $\polQ_2$
finite elements on a grid with refinement level $r=4$ amounting to a total
of $53M$ degrees of freedom; see Section~\ref{sec:numerics_boundary}. In
addition we run the simulation on a second modified grid where the
distances to the upper, lower and side walls has been enlarged by a factor
two. We again set the stabilization parameters to $c_1=2$, and $c_2=1$.
\begin{figure}[t]
  \centering
  \begin{subfigure}[b]{0.9\textwidth}
      \centering
      \includegraphics[trim=0 0 0 0, clip, width=0.9\textwidth]{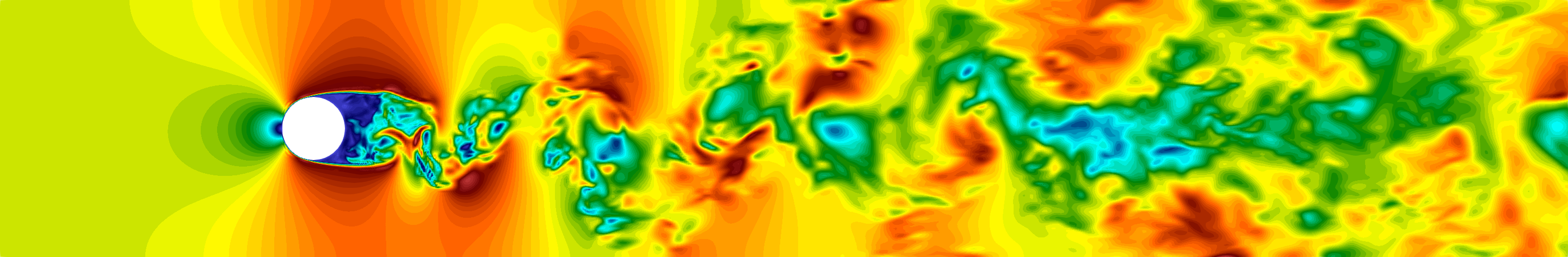}
      \caption{$Re=3900$, narrow domain}
  \end{subfigure}\hfill
  \begin{subfigure}[b]{0.9\textwidth}
      \centering
      \includegraphics[trim=0 0 0 0, clip, width=0.9\textwidth]{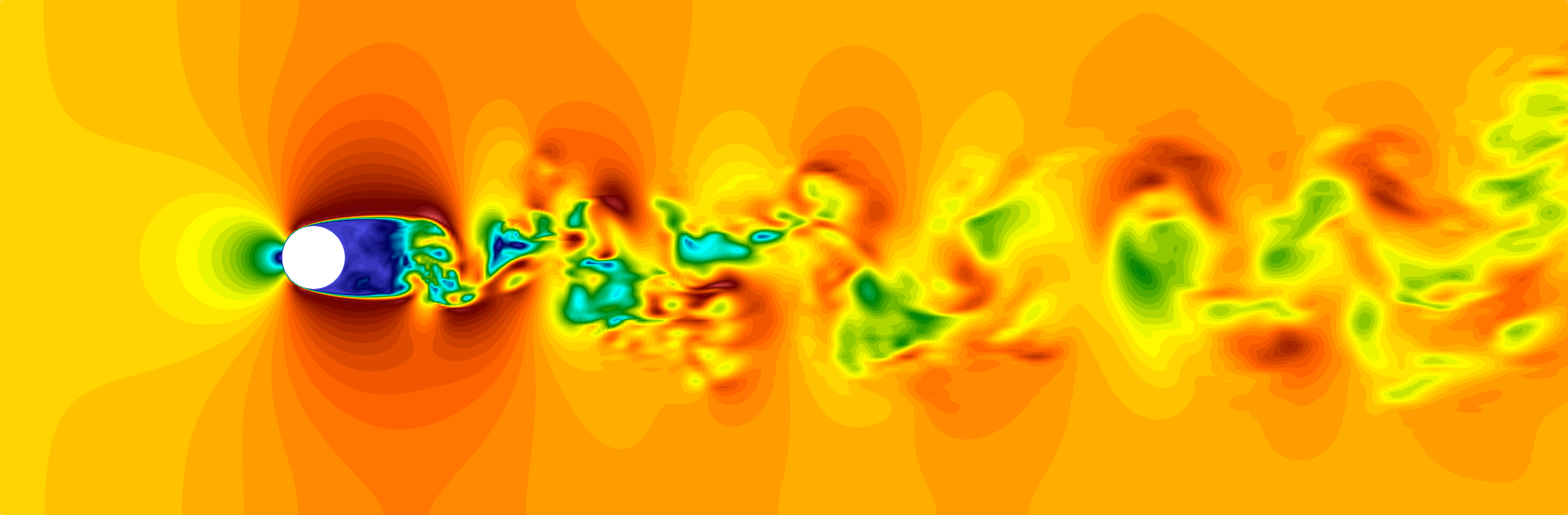}
      \caption{$Re=3900$, wide domain}
  \end{subfigure}\hfill

  \caption{Validation: flow past a 3D cylinder at $Re=3900$ with no-slip
    boundary conditions. 2D cutout (with $z=0$) of a temporal snapshot of
    the velocity magnitudes at time $T = 0.5$ are shown. A no-slip boundary
    condition is applied on the cylinder surface and slip boundary
    conditions are applied on the channel walls. The simulation is
    performed using $\polQ_2$ finite elements with a total of $53M$ degrees
    of freedom.}
  \label{cyl:Hoffman_Re_3900}
\end{figure}
A temporal snapshot of velocity magnitudes at time $t=0.5$ is shown in
Figure~\ref{cyl:Hoffman_Re_3900}. The drag coefficient $c_d$ has been
computed by averaging from on the time interval $[0.2,0.5]$:
\begin{align*}
  \begin{aligned}
    \text{Narrow domain:}\quad
    \mu_{c_d} \,&=\, \mynum{1.34212}, &\quad \sigma_{c_d} \,&=\,
    \mynum{0.0176779},
    \\[0.25em]
    \text{Wide domain:}\quad
    \mu_{c_d} \,&=\, \mynum{1.05577}, &\quad \sigma_{c_d} \,&=\,
    \mynum{0.0105935}.
  \end{aligned}
\end{align*}
The drag value we obtained for the larger domain is in good agreement with
those reported from computational and experimental
results~\citep{hoffman2009}. The drag coefficient for the smaller domain is
slightly larger due to a higher influence of wall effects.

\subsection{Parameter study: surface roughness and flow past a 2D cylinder}
\label{subse:parameter_study_2d}
\begin{figure}[t]
  \centering
  \includegraphics[height=0.65\textwidth]{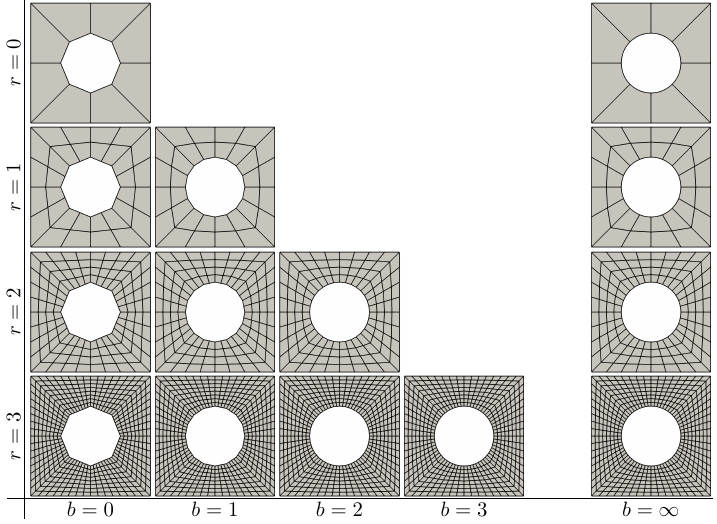}
  \caption{Visualization of the \emph{numerical surface roughness} for
    meshes intoduced into our coputations by refining the coarse mesh $r$
    times. The high-order manifold at the cylinder is removed after $b$
    refinement steps, resulting in a linear $\polQ_1$ mapping. For case
    $b=\infty$ the high-order manifold has not been removed and an
    isoparametric $\polQ_2$ mapping is used.}
    \label{fig:roughness}
\end{figure}
\begin{figure}[t!]
  \centering
  \begin{subfigure}[b]{0.3\textwidth}
      \centering
      \includegraphics[trim=0 0 750 0, clip, width=0.9\textwidth]{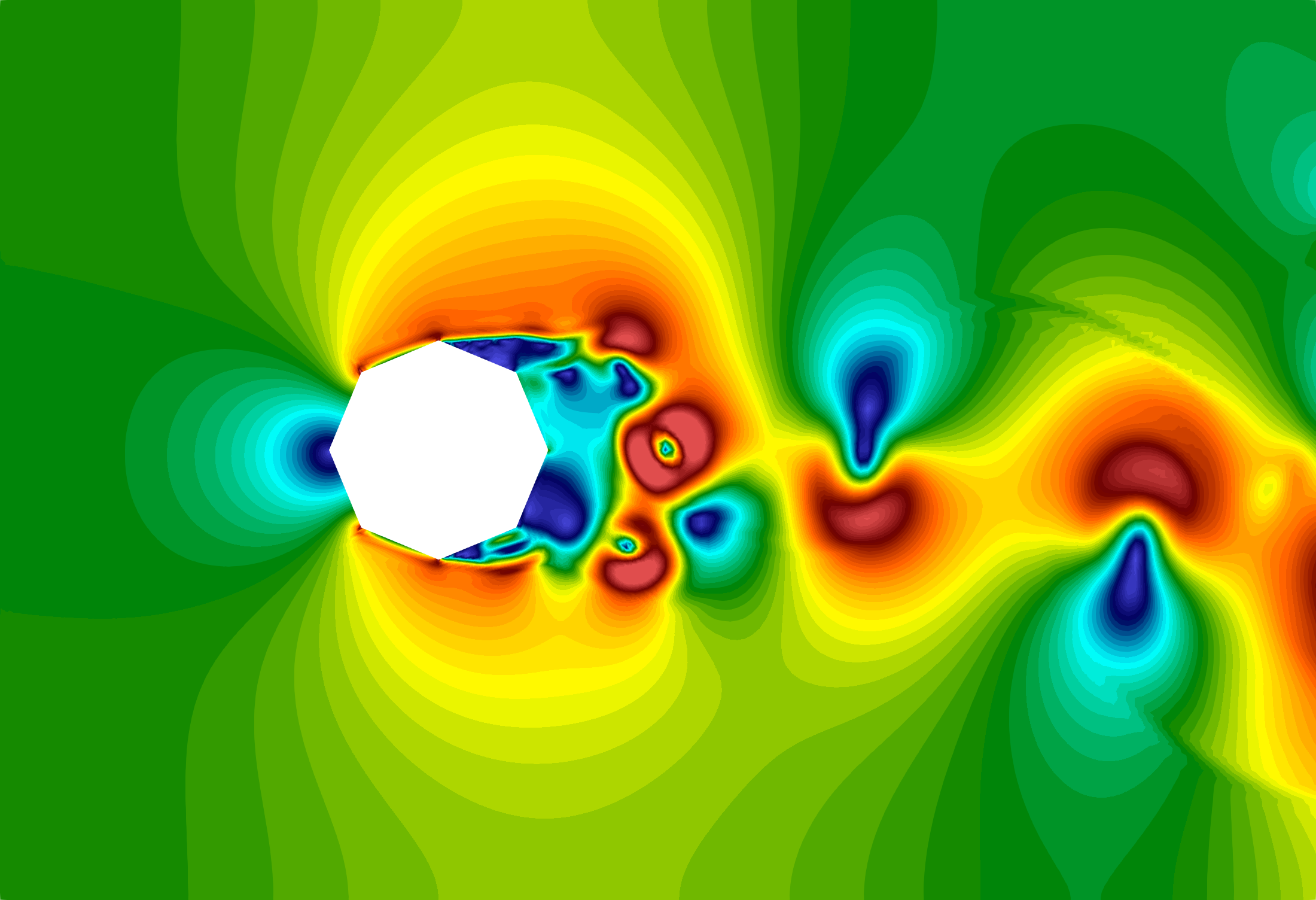}
      \caption{$b=0$ ($d_\text{r}=14.6\,\%$)}
  \end{subfigure}\hfill
  \begin{subfigure}[b]{0.3\textwidth}
      \centering
      \includegraphics[trim=0 0 750 0, clip, width=0.9\textwidth]{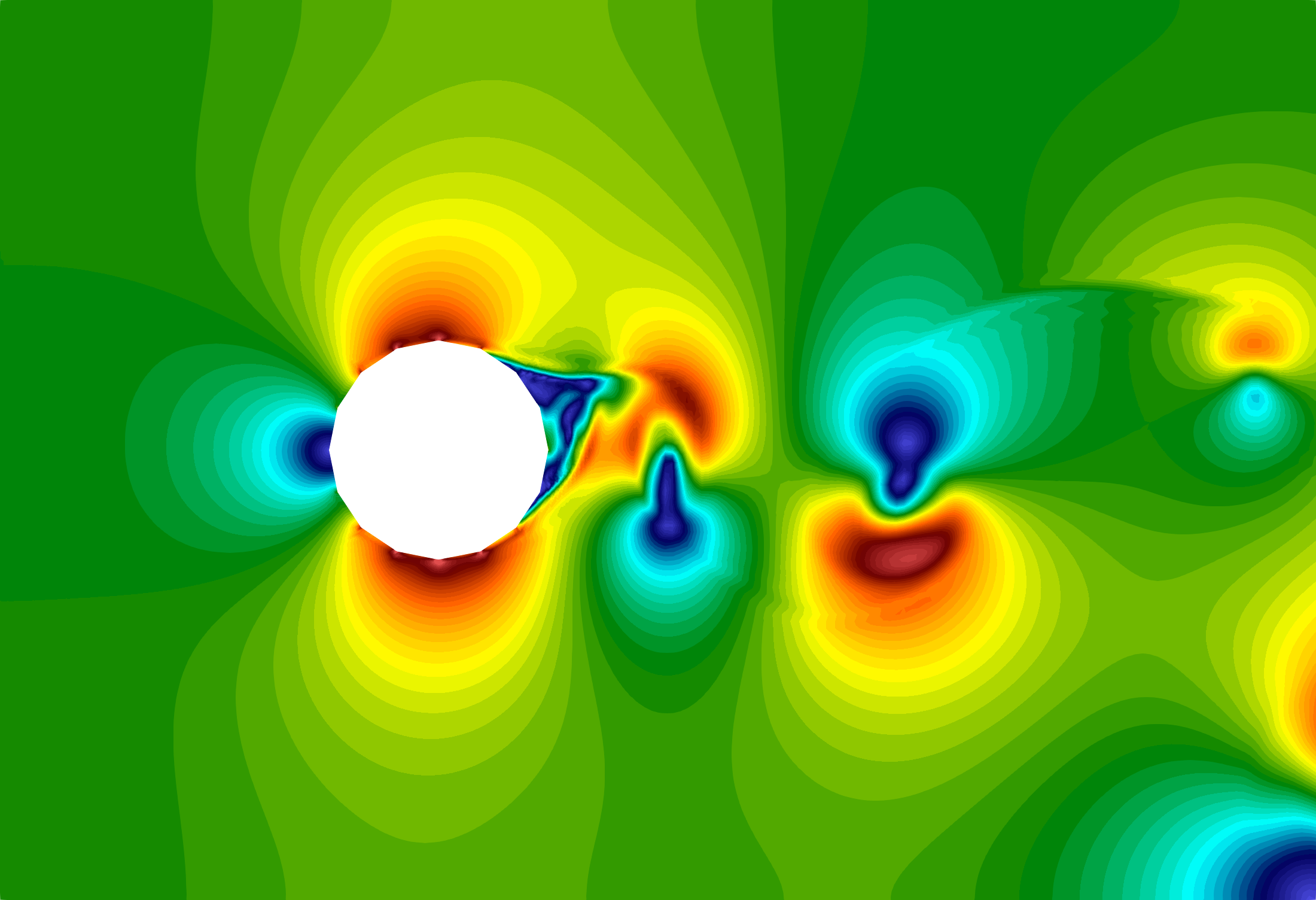}
      \caption{$b=1$ ($d_\text{r}=7.32\,\%$)}
  \end{subfigure}\hfill
  \begin{subfigure}[b]{0.3\textwidth}
      \centering
      \includegraphics[trim=0 0 750 0, clip, width=0.9\textwidth]{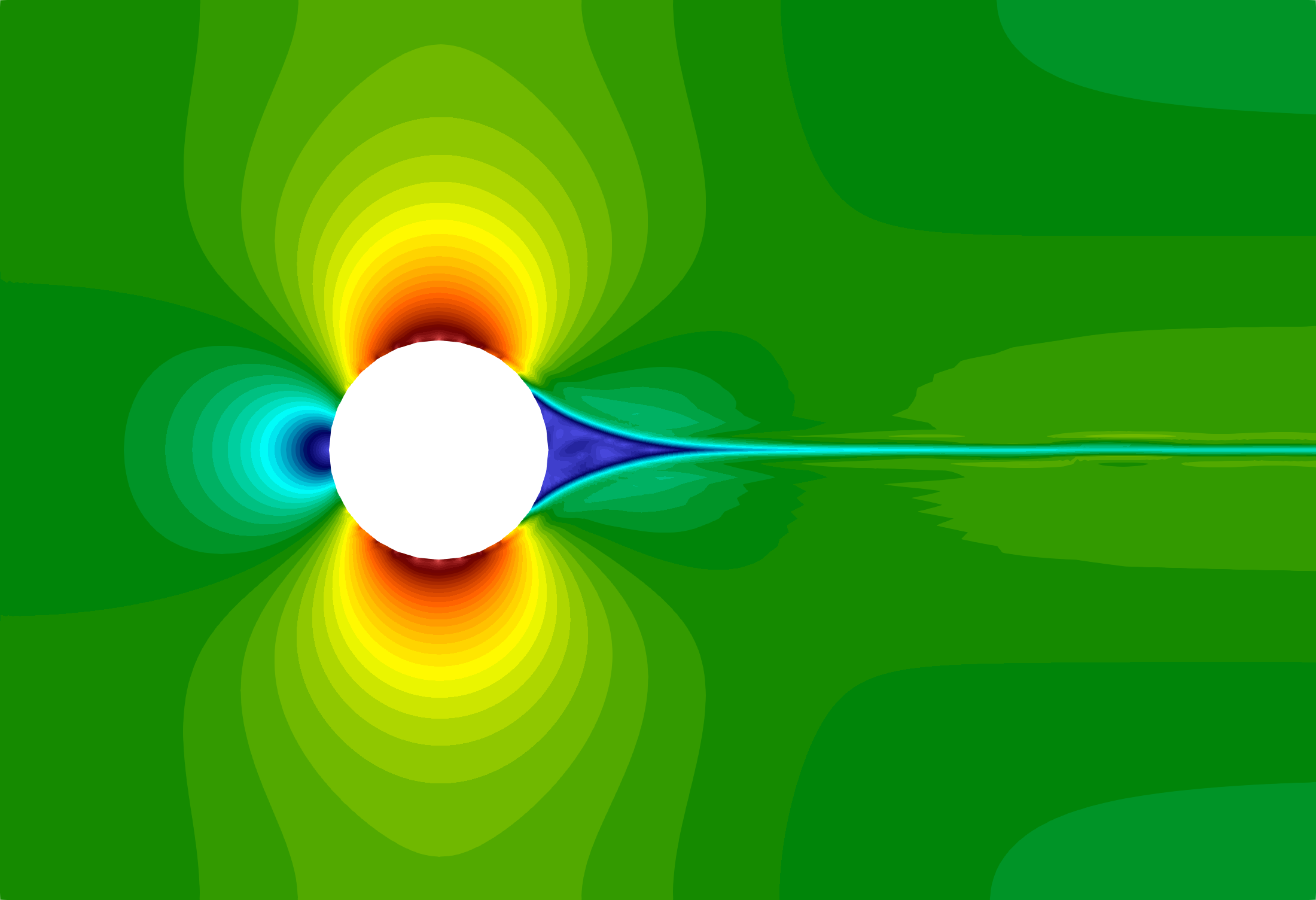}
      \caption{$b=2$ ($d_\text{r}=3.66\,\%$)}
  \end{subfigure}\hfill
  \begin{subfigure}[b]{0.3\textwidth}
      \centering
      \includegraphics[trim=0 0 750 0, clip, width=0.9\textwidth]{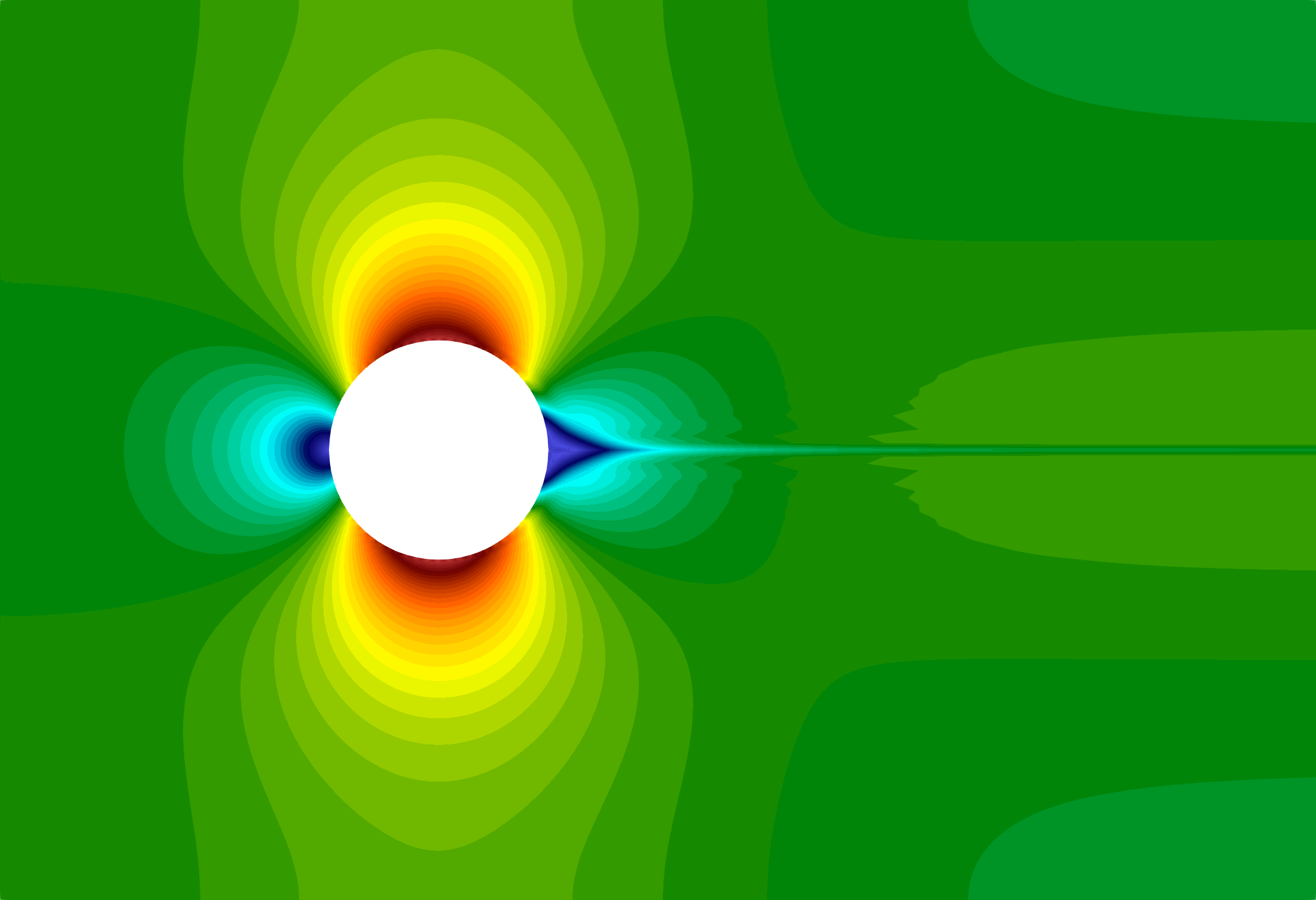}
      \caption{$b=3$ ($d_\text{r}=1.83\,\%$)}
  \end{subfigure}\hfill
  \begin{subfigure}[b]{0.3\textwidth}
      \centering
      \includegraphics[trim=0 0 750 0, clip, width=0.9\textwidth]{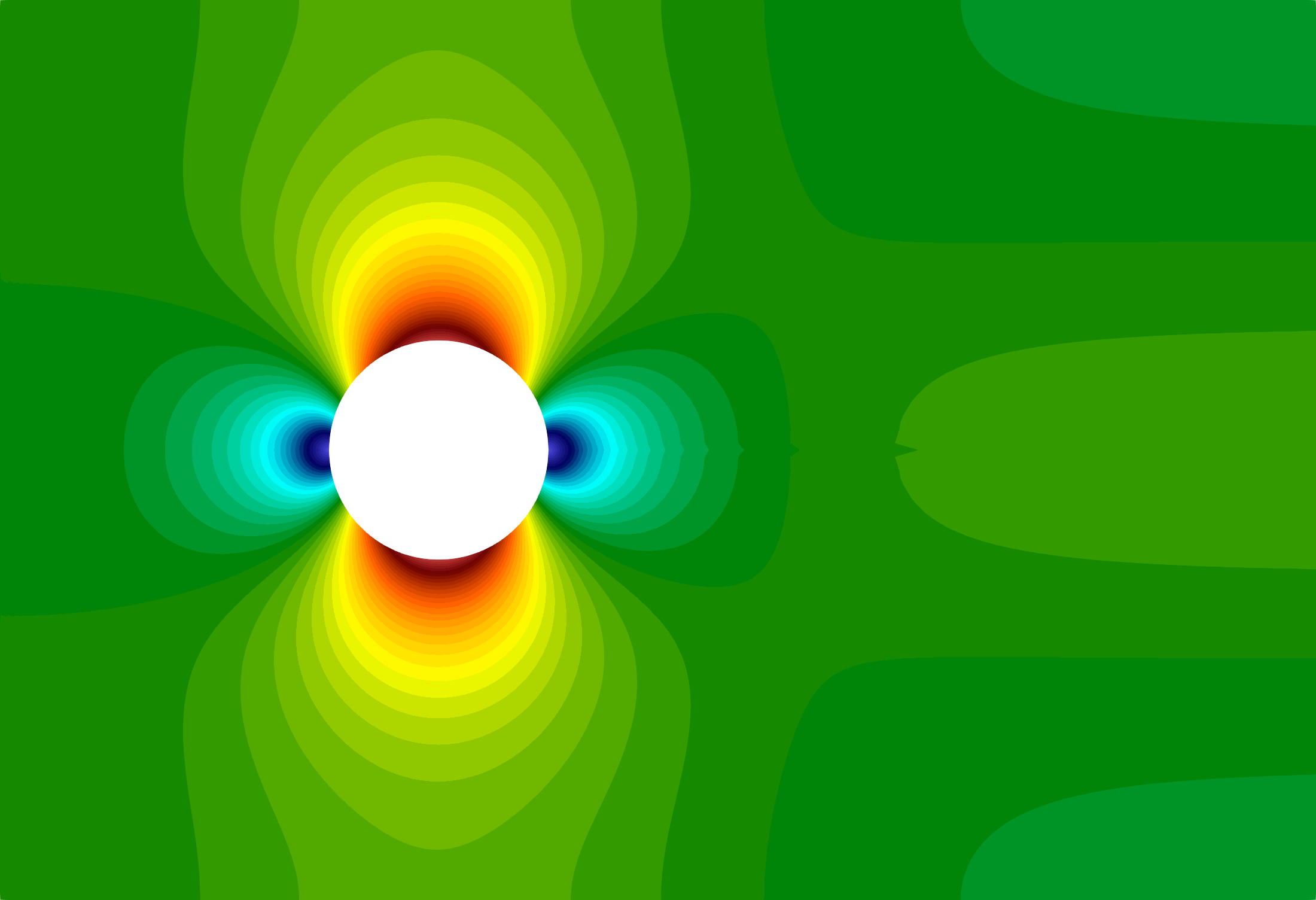}
      \caption{$b=4$ ($d_\text{r}=0.915\,\%$)}
  \end{subfigure}\hfill
  \begin{subfigure}[b]{0.3\textwidth}
      \centering
      \includegraphics[trim=0 0 750 0, clip, width=0.9\textwidth]{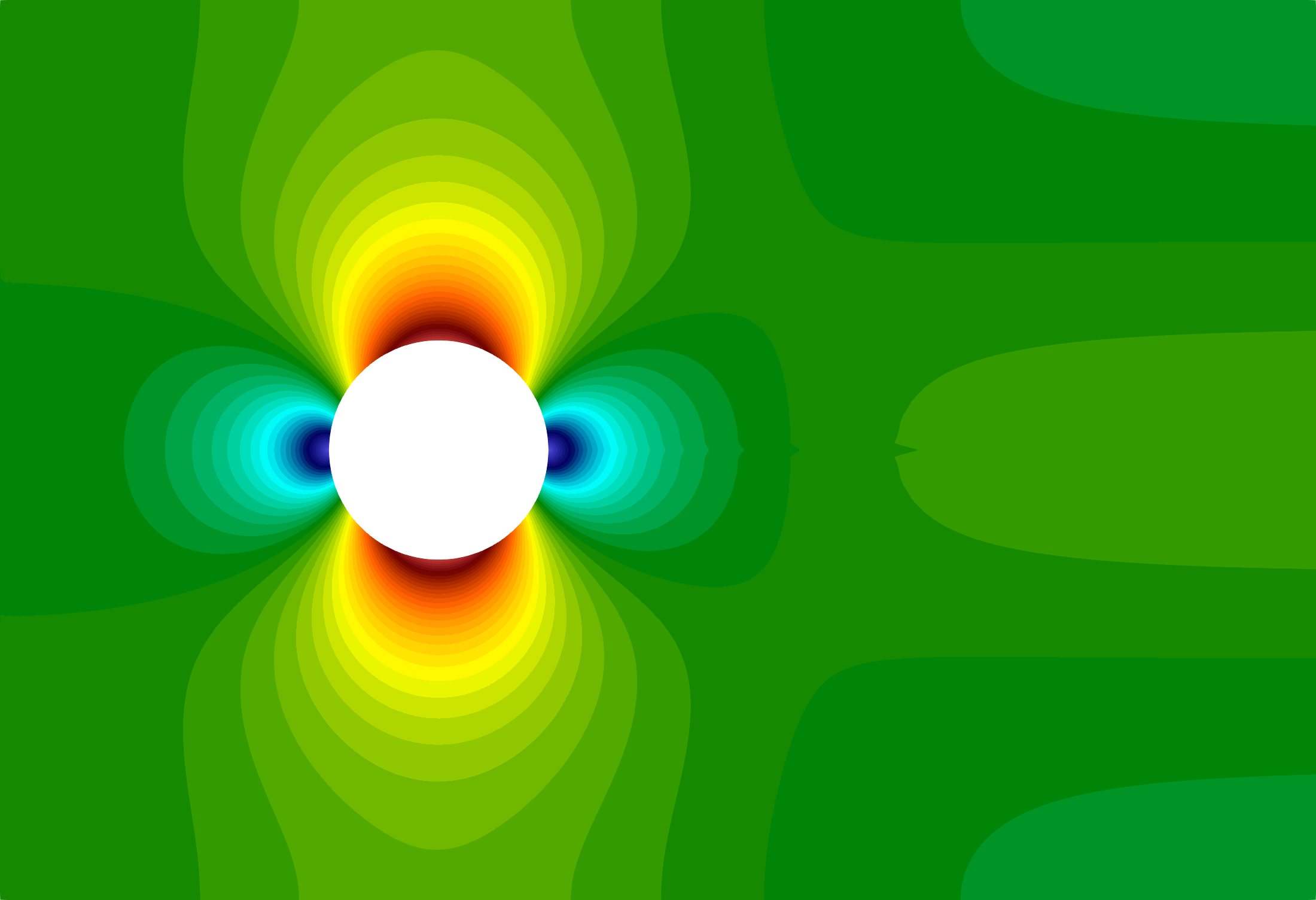}
      \caption{$b=\infty$ ($d_\text{r}\approx0\,\%$)}
  \end{subfigure}\hfill
  \caption{Parameter study with flow past 2D cylinder with continuous
    $\polQ_2$ finite elements on the 2D mesh configuration with refinement
    level $r=4$ amounting to a total number of degrees of freedom of
    $273k$. The high-order $\polQ_2$ manifold on the cylinder is removed
    after $b=0$ to $4$ refinement steps resulting in various different
    degrees of surface roughness. The magnitude of the velocity is shown on
    a rainbow scale in a zoom around the cylinder.}
  \label{fig:2d_u_magnitudes}
  \begin{subfigure}[b]{0.3\textwidth}
      \centering
      \includegraphics[trim=0 0 750 0, clip, width=0.9\textwidth]{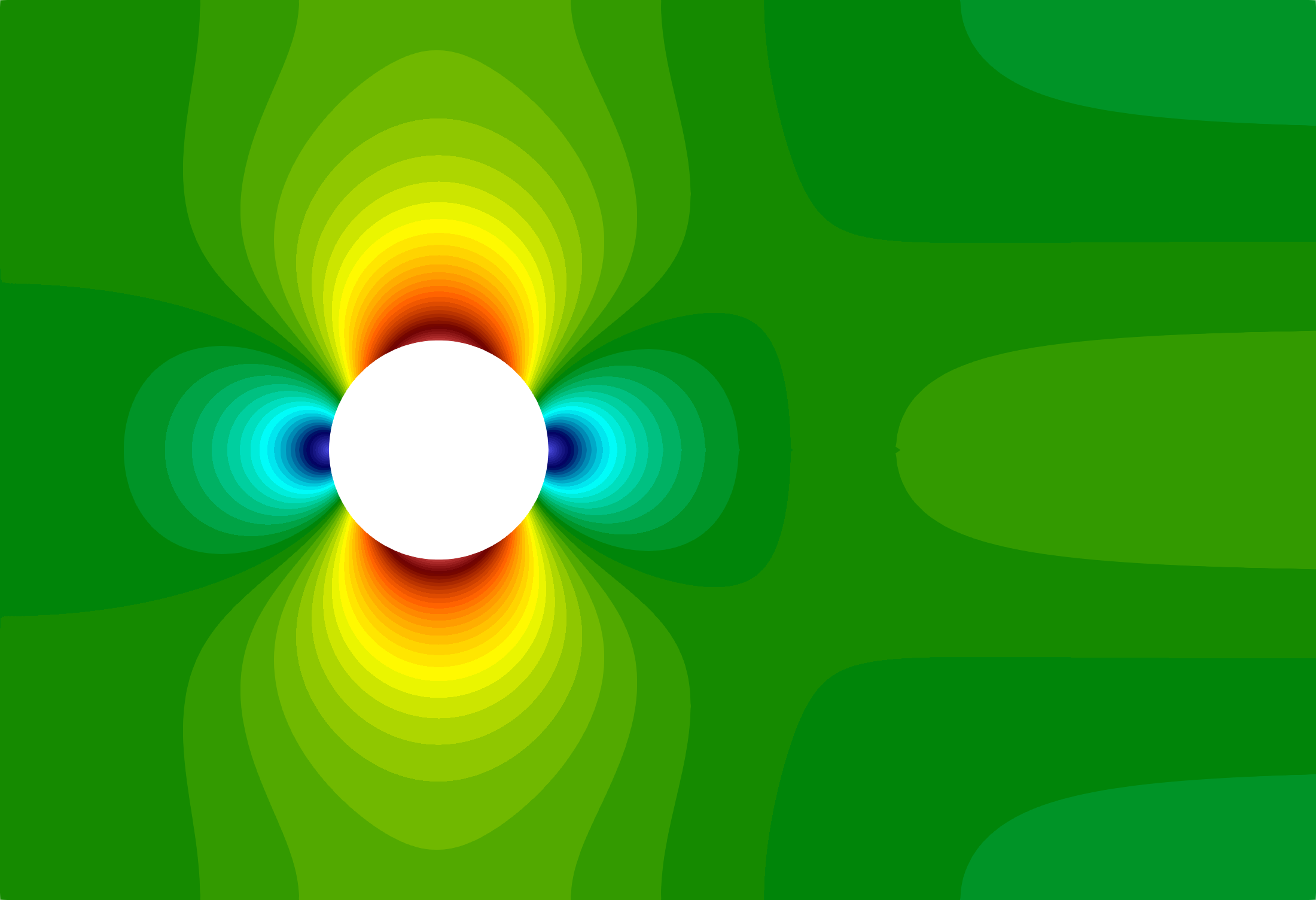}
      \caption{$r=5$, $b=\infty$ ($d_\text{r}\approx0\,\%$)}
  \end{subfigure}\hfill
  \caption{Parameter study with flow past 2D cylinder with continuous
    $\polQ_2$ finite enements on the 2D mesh configuration with refinement
    level $r=5$ amounting to a total number of degrees of freedom of
    $1.092M$. isoparametric $\polQ_2$ approximation ($b=\infty$) of the
    boundary. The magnitude of the velocity is shown on a rainbow scale in
    a zoom around the cylinder.}
  \label{fig:2d_u_magnitudes_r5}
\end{figure}

\begin{table}[t]
  \centering
  \begin{tabular}{lrlrl}
    \toprule
      & $\mu_{c_d}$ & $(\sigma_{c_d})$ & $\mu_{c_l}$ & $(\sigma_{c_l})$
      \\[0.5em]
      $b=0$      & $\num{1.9473}    $ & $(\num{0.386572}   )$ & $\num{0.00367179} $ & $(\num{0.944296}   )$ \\
      $b=1$      & $\num{0.631618}  $ & $(\num{0.162503}   )$ & $\num{0.017183}   $ & $(\num{0.617694}   )$ \\
      $b=2$      & $\num{0.017519}  $ & $(\num{0.00404143} )$ & $\num{-0.00188444}$ & $(\num{0.0413057}  )$ \\
      $b=3$      & $\num{0.00987457}$ & $(\num{3.82538e-07})$ & $\num{5.47641e-11}$ & $(\num{7.6241e-11} )$ \\
      $b=4$      & $\num{0.0075894} $ & $(\num{2.66793e-16})$ & $\num{2.8636e-08} $ & $(\num{2.66717e-22})$ \\
      $b=\infty$ & $\num{0.007622}  $ & $(\num{2.37354e-16})$ & $\num{2.65551e-08}$ & $(\num{6.7615e-22} )$ \\
    \bottomrule
  \end{tabular}\\[1em]

  \begin{tabular}{lrl}
    \toprule
      & $\mu_{\Delta p}$ & $(\sigma_{\Delta p})$
      \\[0.5em]
      $b=0$      & $\num{5142.59}$ & $(\num{2361.49}    )$ \\
      $b=1$      & $\num{2438.41}$ & $(\num{910.42}     )$ \\
      $b=2$      & $\num{516.845}$ & $(\num{25.9809}    )$ \\
      $b=3$      & $\num{204.645}$ & $(\num{0.00202621} )$ \\
      $b=4$      & $\num{7.35176}$ & $(\num{6.90841e-15})$ \\
      $b=\infty$ & $\num{7.45528}$ & $(\num{1.07394e-13})$ \\
    \bottomrule
  \end{tabular}
  \hspace{3em}
  \begin{tabular}{lc}
    \toprule
      & $d_{\text{r}}$
      \\[0.5em]
      $b=0$      & $14.6\,\%$ \\
      $b=1$      & $7.32\,\%$ \\
      $b=2$      & $3.66\,\%$ \\
      $b=3$      & $1.83\,\%$ \\
      $b=4$      & $0.915\,\%$ \\
      $b=\infty$ & $\approx0\,\%$ \\
    \bottomrule
  \end{tabular}
  \caption{2D parameter study: quantities of interest as a function of
    surface roughness $b$ for refinement level $r=4$. Temporal average and
    standard deviation of drag and lift coefficients, $c_d$, $c_l$, and
    pressure difference $\Delta p$ are computed on the time interval
    $[0.2,0.5]$.}
  \label{table:parameter_2d}
\end{table}
We now modify the previous flow configuration as follows. First, as
suggested in \citep[Sec.~9]{Hoffman_2010} we set $\nu=0$ and set \emph{slip
boundary conditions} also on the cylinder. We keep the uniform inflow
profile with $\bar u = u_{\text{max}}=39$, as well as zero initial
conditions and the linear ramp-up of inflow conditions from $t=0$ to
$t=0.05$. We again use continuous $\polQ_2$ finite elements for the 2D mesh
at refinement level $r=4$ ($273k$ total degrees of freedom). In addition to
the isoparametric approximation ($b=\infty$) various degrees of surface
roughness are simulated ($b=0,\ldots, 4$). This is achieved by first
refining the mesh with a high-order manifold description on the cylinder
$b$ times, then removing the manifold description for the remaining $r-b$
refinements, resulting in a linear $\polQ_1$ mapping; see
Section~\ref{sec:numerics_boundary}. The process is visualized in
Figure~\ref{fig:roughness}. As a convenient metric for the surface
roughness, we also report the maximal relative radial deviation
$d_{\text{r}}$ of the discretized boundary from the true circle. The
deviation starts at $14.6\,\%$ for $b=0$, and halfs with every boundary
approximation step.

A temporal snapshot showing velocity magnitudes in a zoom around the
cylinder for time $t=0.5$ are reported in Figure~\ref{fig:2d_u_magnitudes}.
For the simulations we varied the degree of surface roughness from $b=0$ to
$b=4$ and show an isoparametric approximation with $b=\infty$.  We perform
statistics for drag and lift coefficients and the pressure drop on the time
interval $[0.2,0.5]$. We have verified that all simulations are past the
transient startup and have reached a fully developed flow regime at
$t=0.2$. The results are reported in Table~\ref{table:parameter_2d}.
Judging from Figure~\ref{fig:2d_u_magnitudes} it is evident that the flow
becomes stable for decreasing surface roughness while rapidly approaching a
potential flow profile. This is corroborated by the statistics summarized
in Table~\ref{table:parameter_2d}, where, depending on the level of surface
roughness ($b=0$, \ldots $4$, $\infty$), the resulting drag and lift
coefficients vary by 3 to 5 orders of magnitude.

The snapshot of the solution at the final time $t=0.5$ an isoparametric
discretization with a refinement level of $r=5$ ($1.092M$ total degrees of
freedom) are shown in Figure~\ref{fig:2d_u_magnitudes_r5}. For this
resolution, we observe that geometric errors from the cylinder boundary are
almost entirely eliminated. As a result, no boundary error propagates
downstream, and the computed drag force is nearly zero.

\subsection{Parameter study: surface roughness and flow past a 3D cylinder}
\label{subse:parameter_study_3d}
\begin{figure}[t]
  \centering
  \begin{subfigure}[b]{0.5\textwidth}
      \centering
      \includegraphics[trim=0 0 0 0, clip, width=0.9\textwidth]{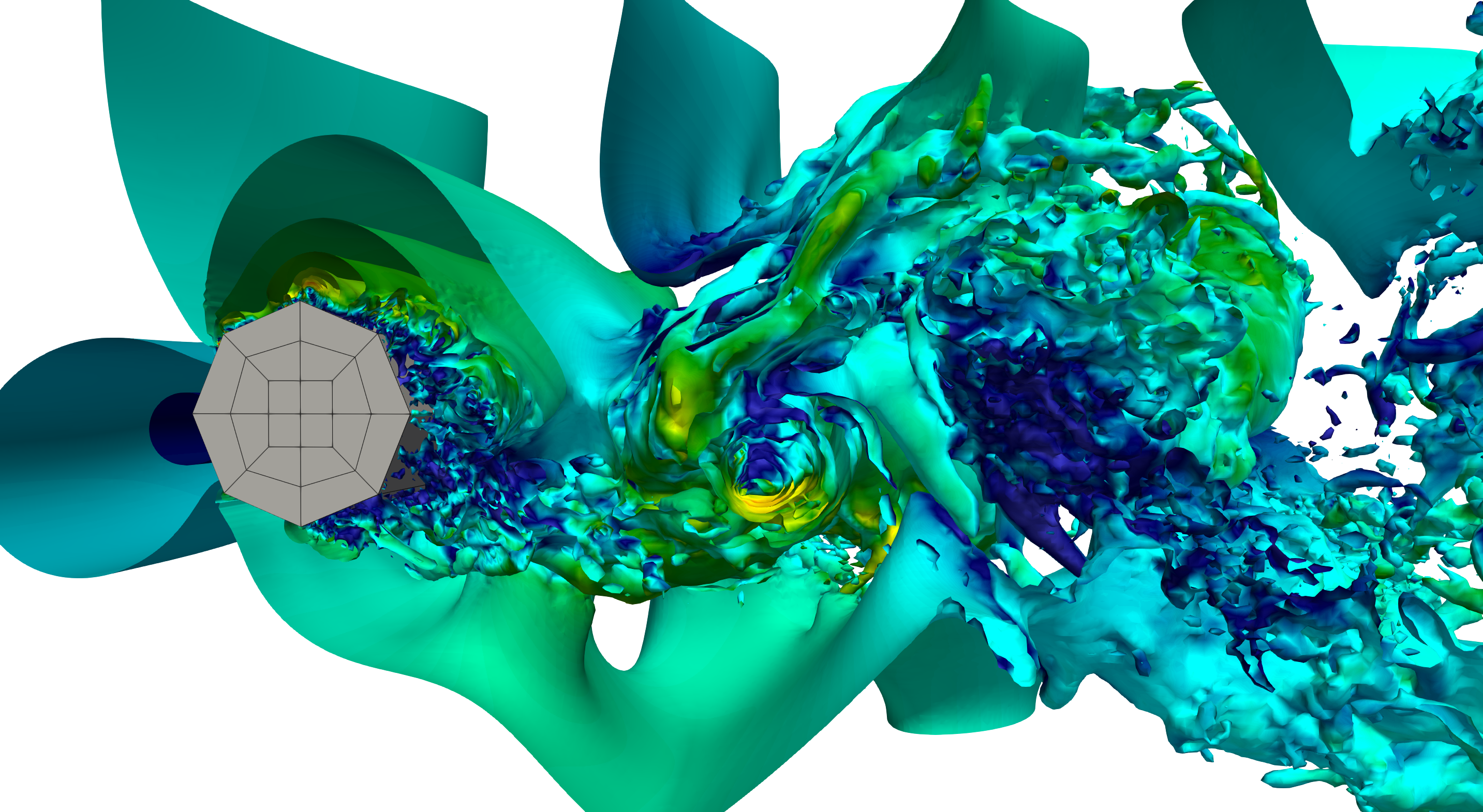}
      \caption{$b=0$ ($d_\text{r}=14.6\,\%$)}
  \end{subfigure}\hfill
  \begin{subfigure}[b]{0.5\textwidth}
      \centering
      \includegraphics[trim=0 0 0 0, clip, width=0.9\textwidth]{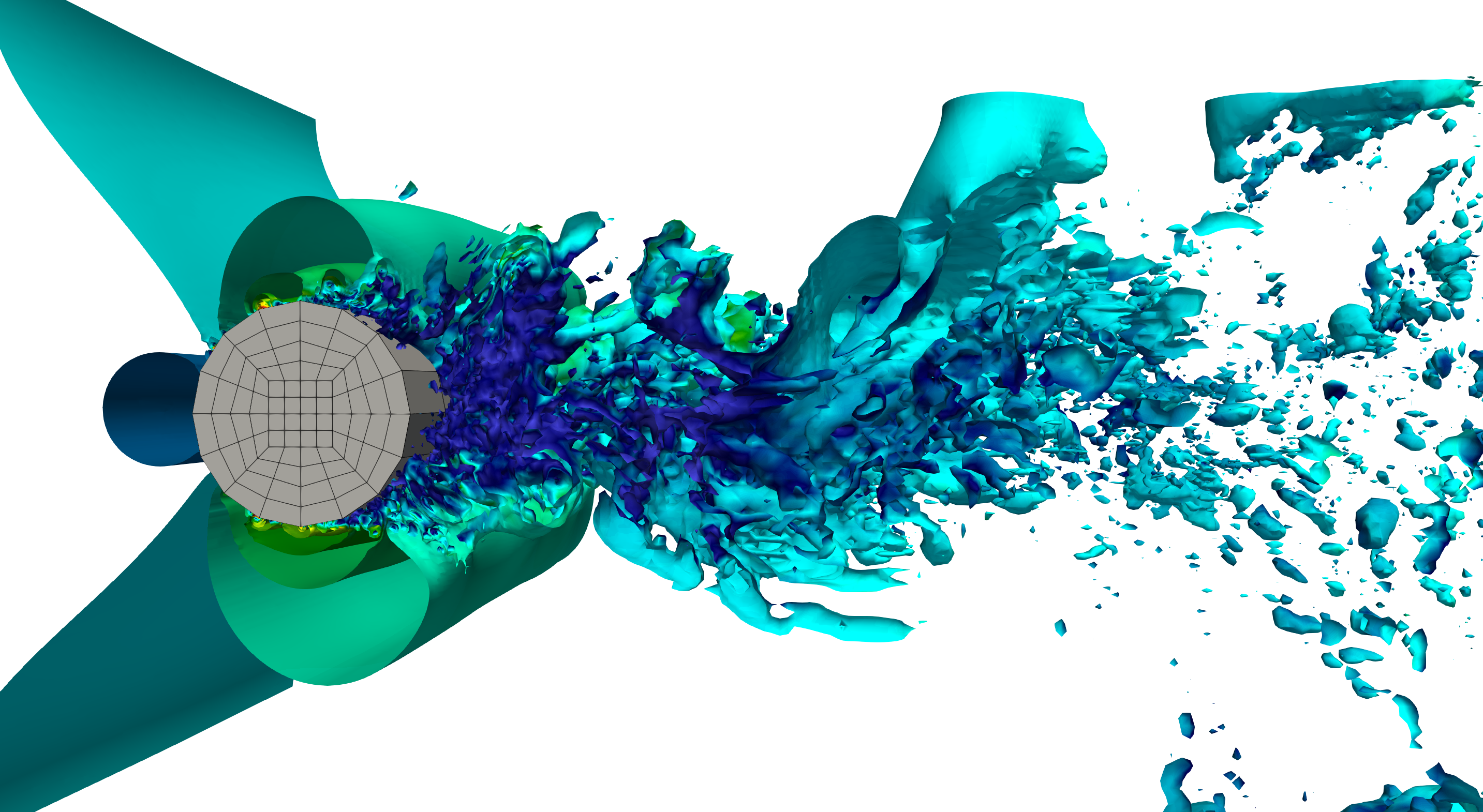}
      \caption{$b=1$ ($d_\text{r}=7.32\,\%$)}
  \end{subfigure}\hfill
  \begin{subfigure}[b]{0.5\textwidth}
      \centering
      \includegraphics[trim=0 0 0 0, clip, width=0.9\textwidth]{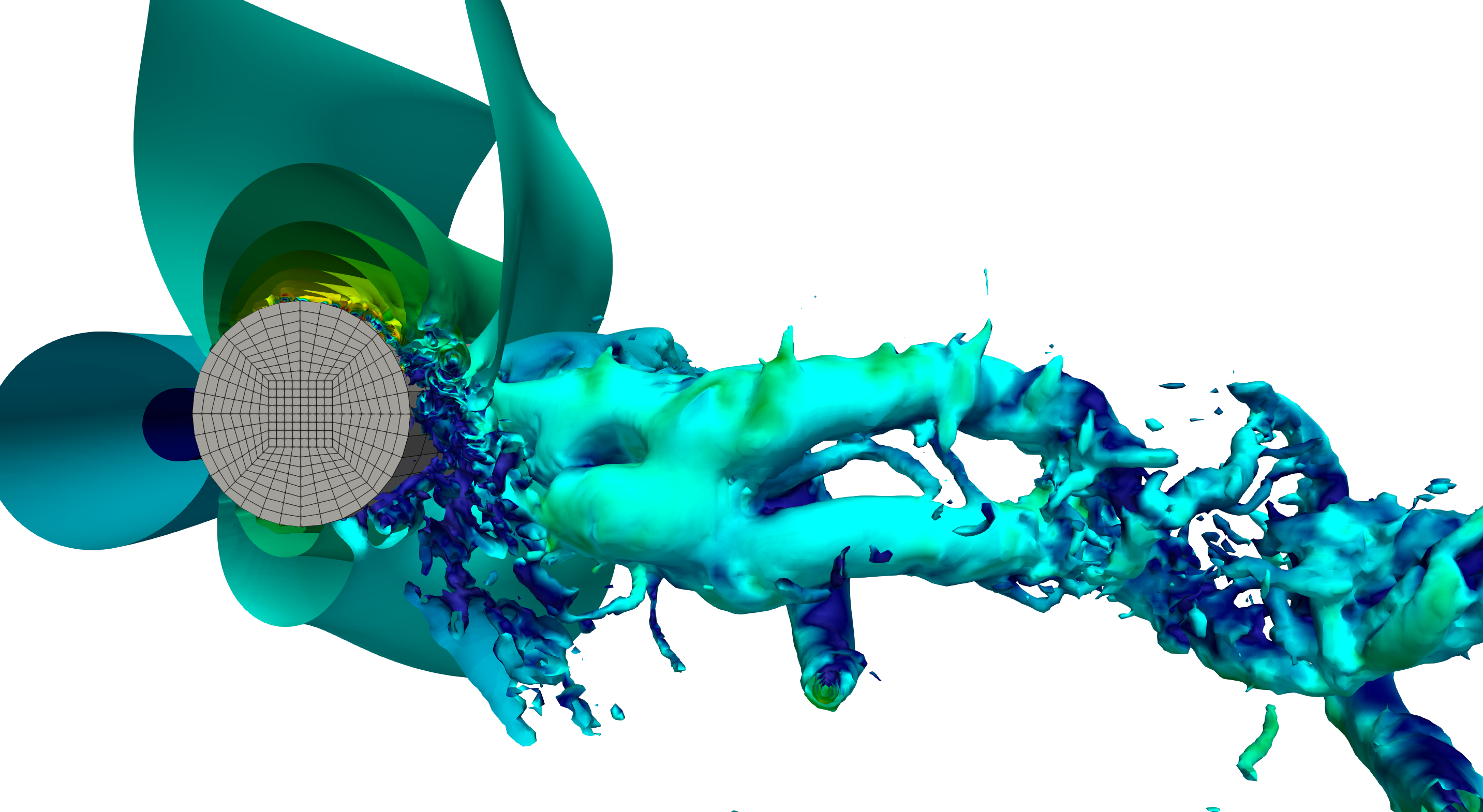}
      \caption{$b=2$ ($d_\text{r}=3.66\,\%$)}
  \end{subfigure}\hfill
  \begin{subfigure}[b]{0.5\textwidth}
      \centering
      \includegraphics[trim=0 0 0 0, clip, width=0.9\textwidth]{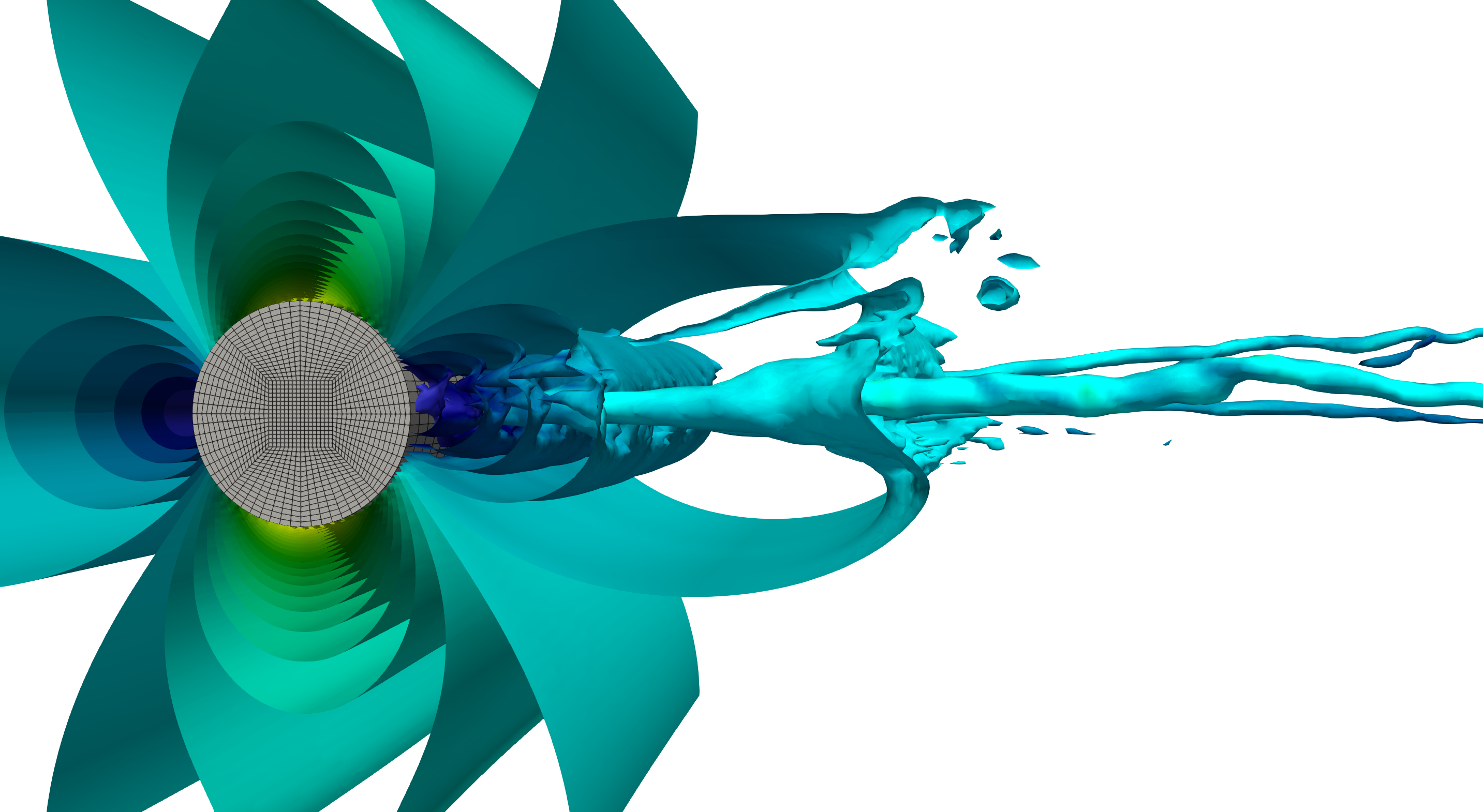}
      \caption{$b=3$ ($d_\text{r}=1.83\,\%$)}
  \end{subfigure}\hfill
  \begin{subfigure}[b]{0.5\textwidth}
      \centering
      \includegraphics[trim=0 0 0 0, clip, width=0.9\textwidth]{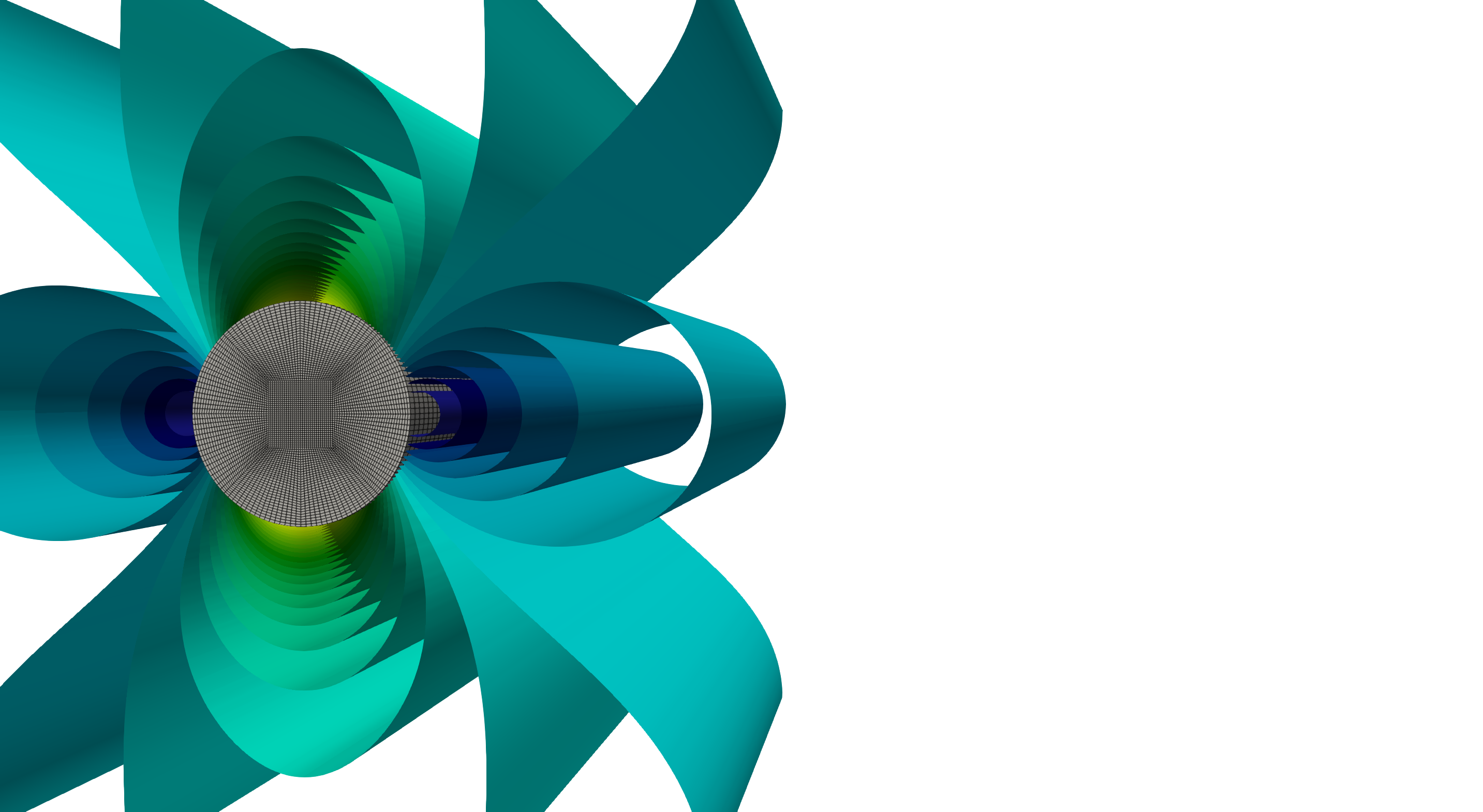}
      \caption{$b=4$ ($d_\text{r}=0.915\,\%$)}
  \end{subfigure}\hfill
  \begin{subfigure}[b]{0.5\textwidth}
      \centering
      \includegraphics[trim=0 0 0 0, clip, width=0.9\textwidth]{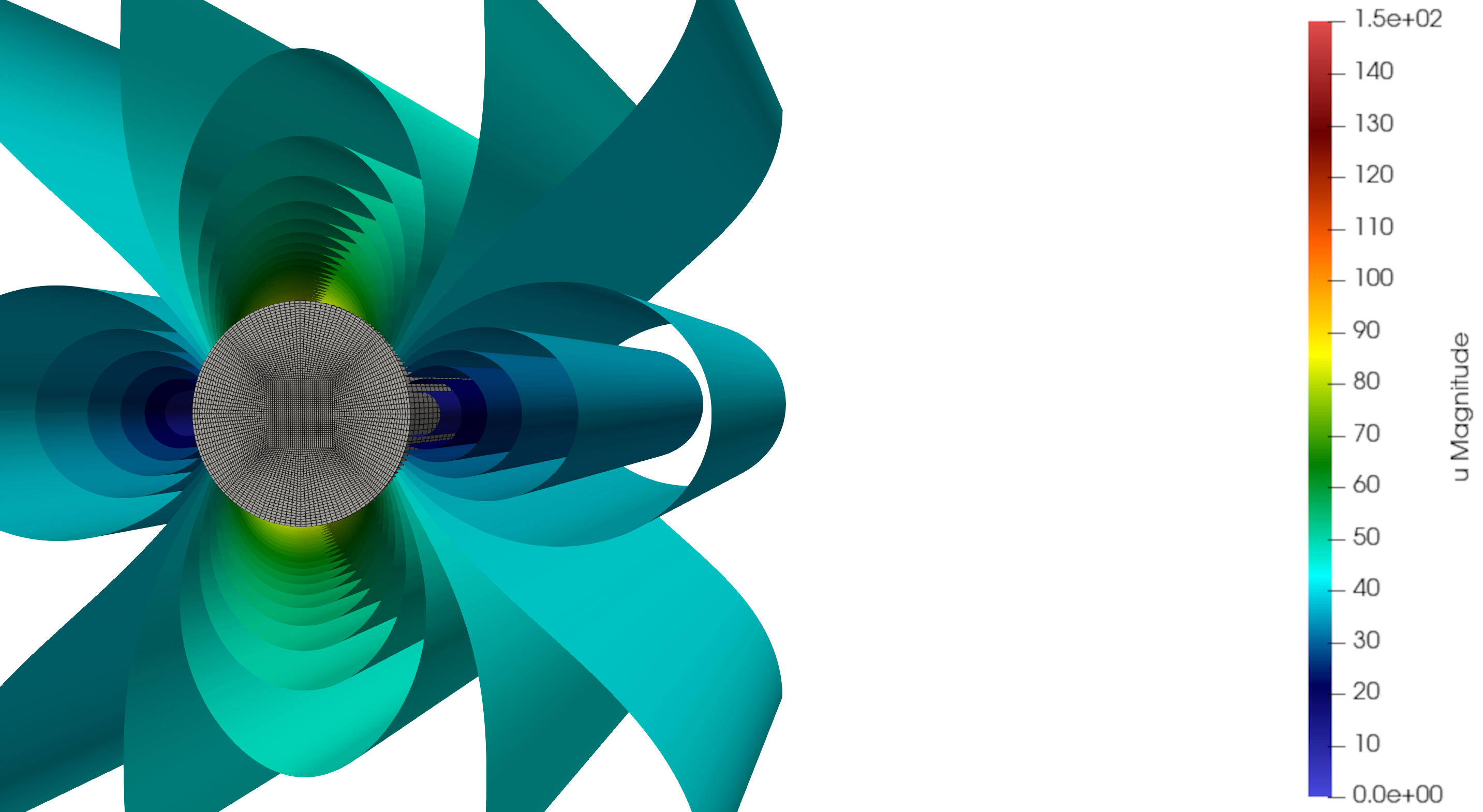}
      \caption{$b=\infty$ ($d_\text{r}\approx0\,\%$)}
  \end{subfigure}\hfill
  \caption{Parameter study with flow past 3D cylinder with continuous
    $\polQ_2$ finite elements on the 3D mesh configuration with refinement
    level is $r=4$ amounting to a total number of degrees of freedom of
    $53M$. The high-order $\polQ_2$ manifold on the cylinder is removed after
    $b=0$ to $4$ refinement steps resulting in various different degrees of
    surface roughness. This is visualized with a gray cylindrical inset
    showing the corresponding coarse grid; see the discussion in
    Section~\ref{subse:parameter_study_2d} and Figure~\ref{fig:roughness}.
    The snapshots show 30 equidistant pressure iso-surfaces between minimal
    and maximal pressure and are colorized with the velocity magnitude on a
    rainbow color scale between $[0,150]$.}
    \label{fig:3d_iso_surfaces}
\end{figure}
\begin{table}[t]
  \centering
  \begin{tabular}{lrlrl}
    \toprule
      & $\mu_{c_d}$ & $(\sigma_{c_d})$ & $\mu_{c_l}$ & $(\sigma_{c_l})$
      \\[0.5em]
      $b=0$      & $\num{1.44406}$      & $(\num{0.0569284})$   & $\num{0.0127182}$    & $(\num{0.533717})$   \\
      $b=1$      & $\num{0.421059}$     & $(\num{0.025631})$    & $\num{-0.0031076}$   & $(\num{0.111055})$   \\
      $b=2$      & $\num{0.144491}$     & $(\num{0.0615897})$   & $\num{-0.0693821}$   & $(\num{0.202194})$   \\
      $b=3$      & $\num{0.0134256}$    & $(\num{0.00957513})$  & $\num{-0.0185981}$   & $(\num{0.0276941})$  \\
      $b=4$      & $\num{-0.000116834}$ & $(\num{1.51604e-17})$ & $\num{6.53309e-10}$  & $(\num{5.3665e-23})$ \\
      $b=\infty$ & $\num{-9.96016e-05}$ & $(\num{1.05701e-17})$ & $\num{-2.66934e-10}$ & $(\num{2.18243e-23})$\\
    \bottomrule
  \end{tabular}\\[1em]

  \begin{tabular}{lrl}
    \toprule
      & $\mu_{\Delta p}$ & $(\sigma_{\Delta p})$
      \\[0.5em]
      $b=0$      & $\num{2616.65}$ & $(\num{487.739})$     \\
      $b=1$      & $\num{1165.69}$ & $(\num{64.844})$      \\
      $b=2$      & $\num{436.43}$  & $(\num{30.0896})$     \\
      $b=3$      & $\num{169.022}$ & $(\num{7.26555})$     \\
      $b=4$      & $\num{5.17729}$ & $(\num{4.0006e-13})$  \\
      $b=\infty$ & $\num{5.24576}$ & $(\num{6.50018e-13})$ \\
    \bottomrule
  \end{tabular}
  \hspace{3em}
  \begin{tabular}{lc}
    \toprule
      & $d_{\text{r}}$
      \\[0.5em]
      $b=0$      & $14.6\,\%$ \\
      $b=1$      & $7.32\,\%$ \\
      $b=2$      & $3.66\,\%$ \\
      $b=3$      & $1.83\,\%$ \\
      $b=4$      & $0.915\,\%$ \\
      $b=\infty$ & $\approx0\,\%$ \\
    \bottomrule
  \end{tabular}
  \caption{3D parameter study: quantities of interest as a function of
    surface roughness $b$ for refinement level $r=4$. Temporal average and
    standard deviation of drag and lift coefficients, $c_d$, $c_l$, and
    pressure difference $\Delta p$ are computed on the time interval
    $[0.2,0.5]$.}
  \label{table:parameter_3d}
\end{table}
In light of the stability analysis conducted in~\citep{Hoffman_2010} and
\citep[Chap.~12]{Hoffman_2007}, that establish that the potential flow
solution is unstable at separation points in three spatial dimensions,
resulting to strong vorticity generations in the streamwise direction, the
main question is of course whether the numerical results presented in
Section~\ref{subse:parameter_study_2d} carry over to three space
dimensions. To this end we repeat the parameter study presented previously
but for three spatial dimensions. Except for using the 3D mesh depicted in
Figure~\ref{fig:geometries} we keep all other parameters unchanged. We
choose a refinement level of $r=4$ with $\polQ_2$ finite elements amounting
to a total of $53M$ degrees of freedom.
We set the stabilization parameters to $c_1=1$, and $c_2=1$.

Note that the mesh is symmetric along the $x_2$ and $x_3$ directions and is
uniformly structured. As a result, the GLS stabilization term is
distributed symmetrically and uniformly throughout the domain, avoiding the
introduction of additional numerical artifacts.

Temporal snapshots for time $t=0.5$ depicting pressure iso-surfaces around
the cylinder are shown in Figure~\ref{fig:3d_iso_surfaces}. For the
simulations we varied the degree of surface roughness from $b=0$ to $b=4$
and also report an isoparametric approximation with $b=\infty$. In
Figure~\ref{fig:3d_iso_surfaces} we visualize the surface roughness with a
gray cylindrical inset showing the corresponding coarse grid that defines
the obstacle. We note that all simulations are run at the same global
refinement level of $r=4$.

Similarly to the 2D parameter study, we perform statistics for drag and
lift coefficients and the pressure drop on the time interval $[0.2,0.5]$.
We have verified that all simulations are past the transient startup and
have reached a fully developed flow regime at $t=0.2$. The results are
reported in Table~\ref{table:parameter_3d}. We make the observation that
the flow behaves qualitatively and quantitatively similar to the 2D
parameter study. It is evident that the flow profile becomes stable for
decreasing surface roughness while rapidly approaching a potential flow
profile. This is again confirmed by the statistics summarized in
Table~\ref{table:parameter_3d}, where, depending on the level of surface
roughness ($b=0$ to $4$, $\infty$), the resulting drag and lift
coefficients vary by 3 to 5 orders of magnitude.

\paragraph{Comparison with \cite{Hoffman_2007}}
We conclude the parameter study with a short comparison of our
computational results to \citep{Hoffman_2007} that originally introduced
this benchmark. While studying the benchmark configuration, the authors of
\citep{Hoffman_2007} reported the following observations
\citep[p.~88]{Hoffman_2007}:
\begin{enumerate}
  \item[(i)] Only one separation point is on the downstream of the cylinder;
  \item[(ii)] No boundary layer is prior to the separation point;
  \item[(iii)] Strong generation of streamwise vortices at the separation
    point.
\end{enumerate}
Based on our numerical simulations, we confirm that the first two
observations also hold in the isoparametric case with $b = \infty$.
However, strong streamwise vorticity is observed only when the geometry
error is sufficiently large. For example, in
Figure~\ref{fig:3d_iso_surfaces}(d), where the surface roughness is set to
$b = 3$, significant streamwise vorticity is present. As the surface
roughness decreases, for instance, for $b = 4$ and $b = \infty$
(cf.~Figures~\ref{fig:3d_iso_surfaces}(e) and (f)), no streamwise vortices
or streamwise instabilities are observed.

Finally, we remark in passing that---while not reported in this
manuscript---we observe the same qualitative and quantitative behavior of
the flow for differing refinement levels ($r=3$, $5$, etc.), i.e., for a
given refinement level $r$ we are reliable able to tune the resulting flow
characteristics (drag, lift, pressure drop) 3 to 5 orders of magnitude by
choosing a different boundary roughness $b$.
Similary, we have also verified the behavior of using the inconsistent
stabilization term $S_{\text{G2}}(\bu_h^n, p_h^n ; \bv, q)$ given by
\eqref{eq:stab_inconsistent} instead of the consistent GLS stabilization
term given by \eqref{eq:stab_consistent}. The inconsistent stabilization
gives very similar results which we omit to present.

\subsection{Parameter study: mesh distortion and flow past a 3D cylinder}
\label{subse:parameter_study_3d_mesh_distortion}
\begin{figure}[t]
  \centering
  \begin{subfigure}[b]{0.5\textwidth}
      \centering
      \includegraphics[trim=0 0 0 0, clip,width=0.9\textwidth]{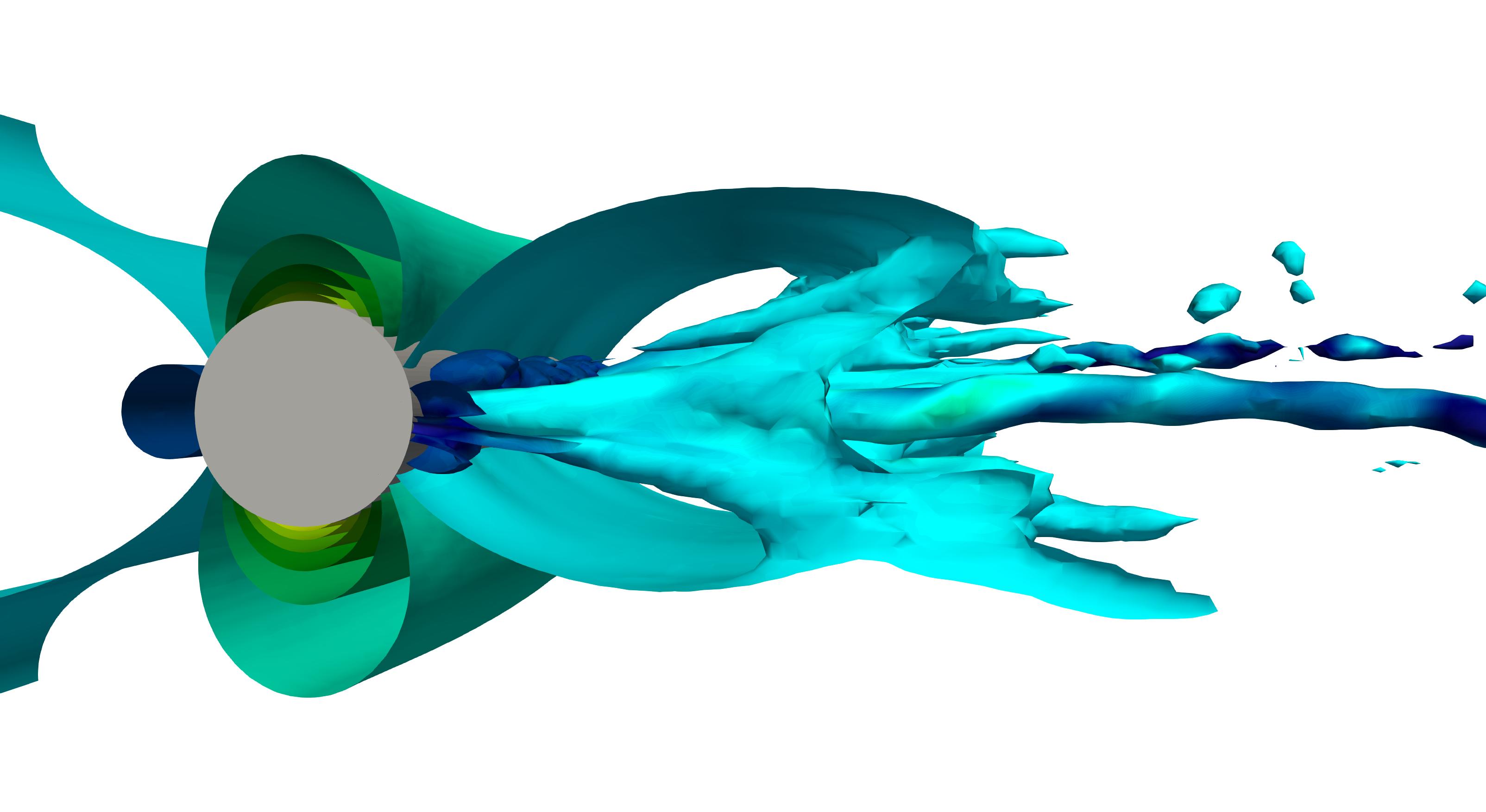}
      \caption{refinement $r=3$, $10\,\%$ perturbation}
  \end{subfigure}\hfill
  \begin{subfigure}[b]{0.5\textwidth}
      \centering
      \includegraphics[trim=0 0 0 0, clip,width=0.9\textwidth]{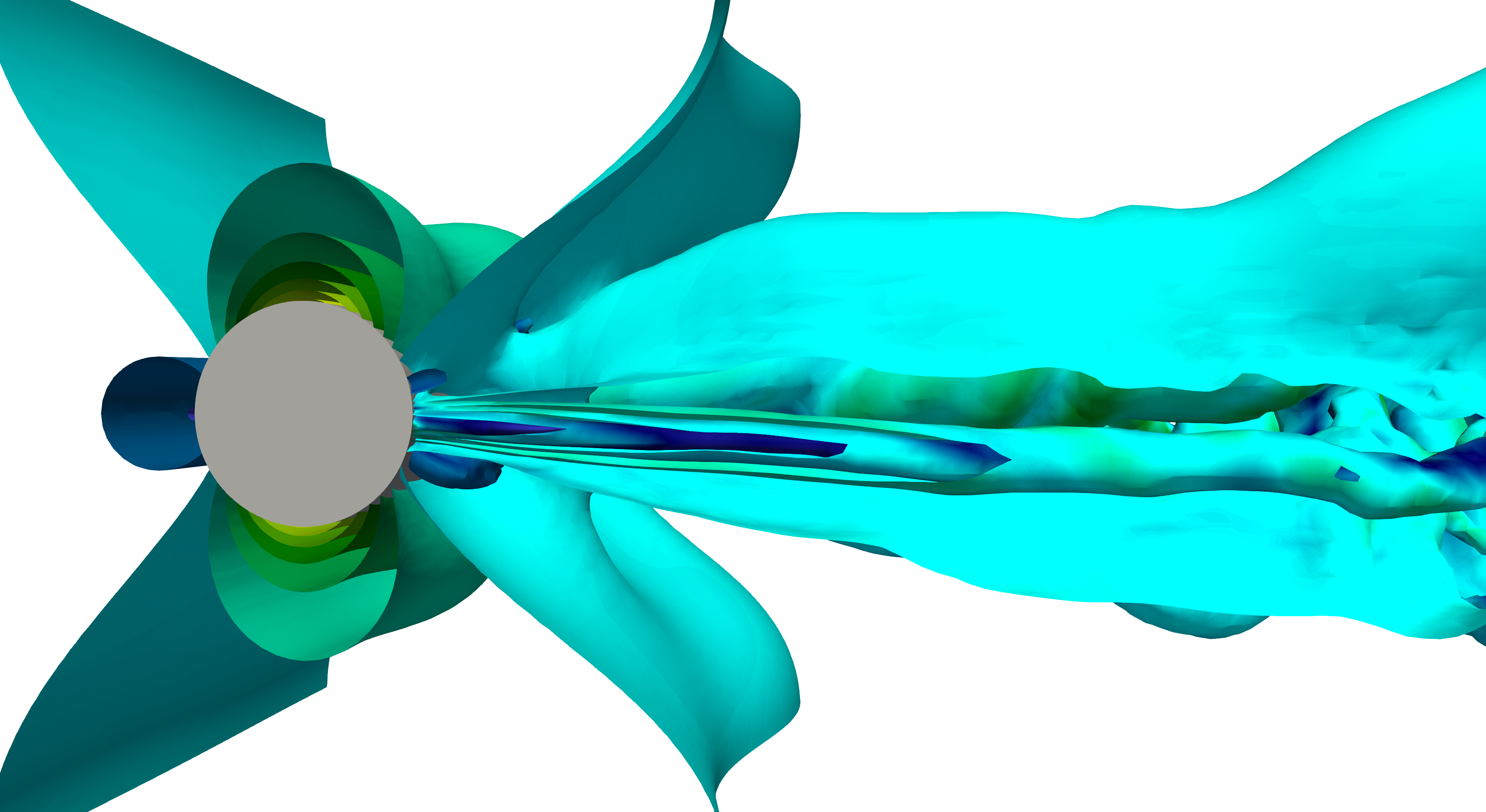}
      \caption{refinement $r=3$, $5\,\%$ perturbation}
  \end{subfigure}\hfill
  \begin{subfigure}[b]{0.5\textwidth}
      \centering
      \includegraphics[trim=0 0 0 0, clip,width=0.9\textwidth]{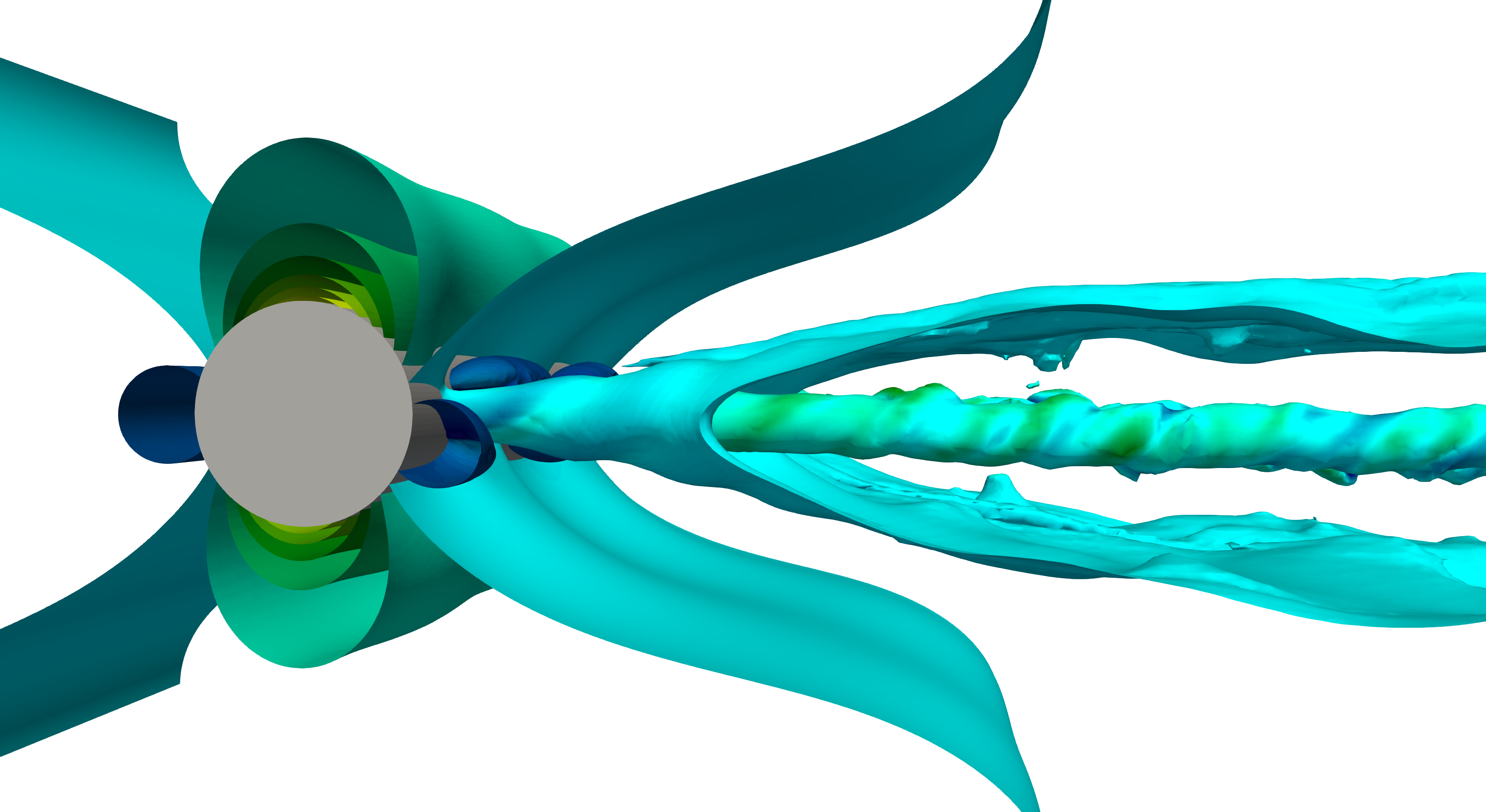}
      \caption{refinement $r=4$, $10\,\%$ perturbation}
  \end{subfigure}\hfill
  \begin{subfigure}[b]{0.5\textwidth}
      \centering
      \includegraphics[trim=0 0 0 0, clip,width=0.9\textwidth]{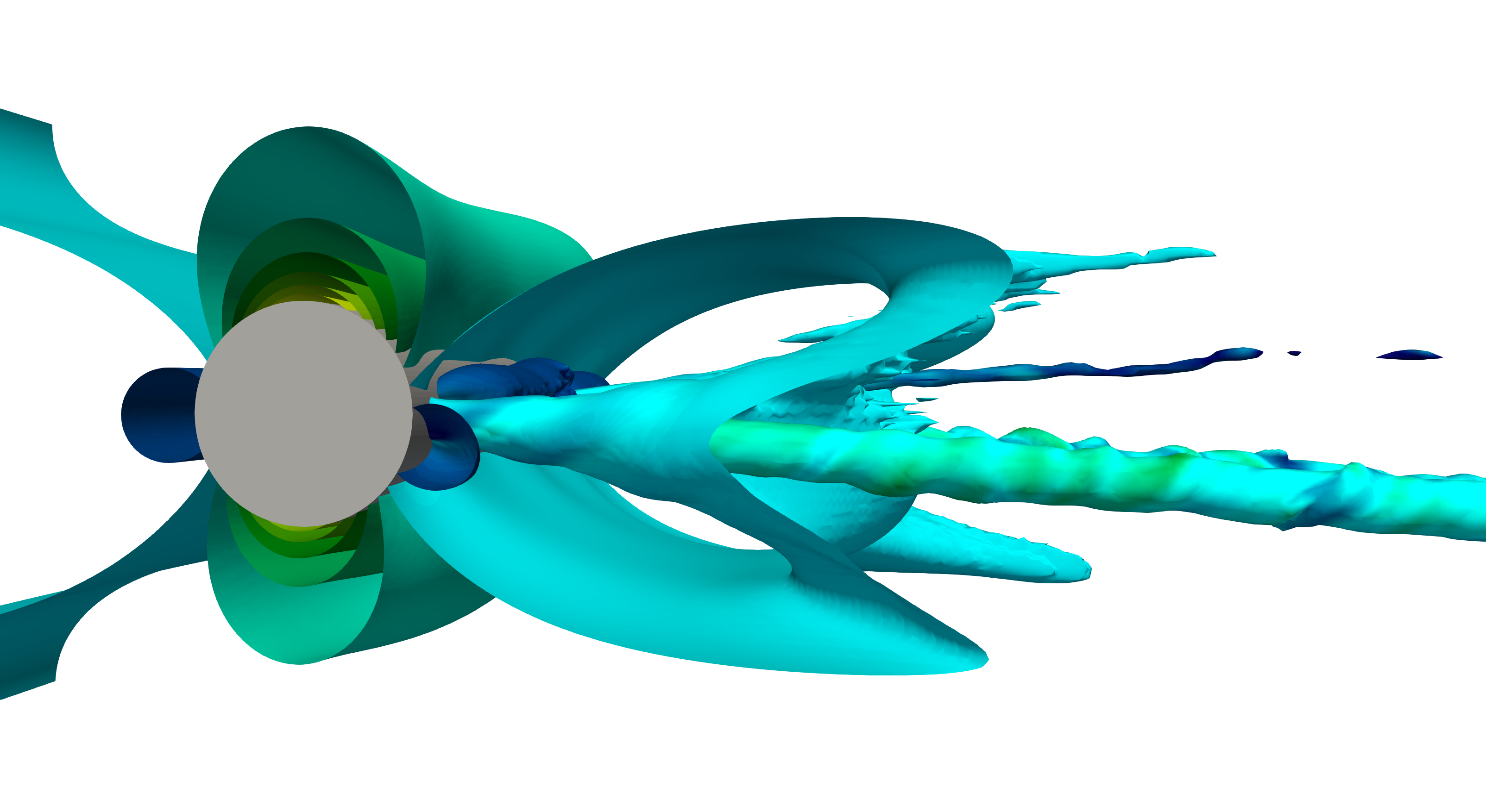}
      \caption{refinement $r=4$, $5\,\%$ perturbation}
  \end{subfigure}\hfill
  \caption{Parameter study with flow past 3D cylinder with continuous
    $\polQ_2$ isoparametric finite elements on the 3D mesh configuration
    with refinement levels $r=3$, $4$ at time $t=1.5$, amounting to a total
    number of degrees of freedom of $6.3M$, and $53M$ respectively. A mesh
    distortion of $10\%$ and $5\%$ is introduced, respectively. The
    snapshots show 30 equidistant pressure iso-surfaces between minimal and
    maximal pressure and are colorized with the velocity magnitude on a
    rainbow color scale between $[0,150]$.}
    \label{fig:3d_iso_surfaces_mesh_distortion}
\end{figure}
\begin{table}
\centering
  \begin{tabular}{ccccc}
    \toprule
                  & \bfseries{10\,\% pert.} & \bfseries{5\,\% pert.} & \bfseries{2.5\,\% pert.} & \bfseries{0\,\% pert.} \\[0.5em]
    $\mathbf r\mathbf=\mathbf2$ & \num{0.59969}     & \num{0.59125}    & \num{0.573252}    & \num{-0.00286983}   \\
                                & (\num{0.162421})  & (\num{0.146364}) & (\num{0.151127})  & (\num{3.06659e-18}) \\[0.25em]
    $\mathbf r\mathbf=\mathbf3$ & \num{0.269627}    & \num{0.44738}    & \num{0.280703}    & \num{-0.000684447}  \\
                                & (\num{0.0710685}) & (\num{0.167142}) & (\num{0.0631349}) & (\num{7.5898e-18})  \\[0.25em]
    $\mathbf r\mathbf=\mathbf4$ & \num{0.13008}     & \num{0.070781}   & \num{0.110479}    & \num{-9.96016e-05}  \\
                                & (\num{0.0466849}) & (\num{0.0151613})& (\num{0.0129083}) & (\num{1.05701e-17}) \\[0.25em]
    \bottomrule
  \end{tabular}
  \caption{3D parameter study: temporal average of the drag coefficient,
    $\mu_{\Delta p}$, and standard deviation $\sigma_{\Delta p}$ as a
    function of refinement level $r=2$, $3$, $4$ and percentage of applied
    random mesh distortion of interior vertices. Temporal average and
    standard deviation are computed on the time interval $[0.5,1.5]$. For
    comparison, we repeat the values obtained for the case of no mesh
    distortion in the last column. Here, statistics are computed over the
    shorter time interval $[0.2,0.5]$.}
  \label{table:parameter_3d_mesh_distortion}
\end{table}
We now turn to the final numerical experiment of this study. The goal of
this test is to investigate whether and how random mesh distortion
influences the flow dynamics of incompressible fluid past a cylinder in
three spatial dimensions. Starting from the 3D mesh with refinement levels
$r=2$, $3$, and $4$, we use again a $\polQ_2$ isoparametric finite element
approximation, resulting in approximately $0.8M$, $6.6M$, $53M$ degrees of
freedom, respectively. We again set the stabilization parameters to
$c_1=1$, and $c_2=1$.

To introduce distortion, every interior vertex of the mesh is perturbed
randomly by up to $p=10\%$, $5\%$, and $2.5\%$ of the minimal distance to
the nearest neighboring vertex. This produces an unstructured mesh in which
the elements exhibit varying volumes. Consequently, the resulting GLS
stabilization terms become nonuniform, breaking the mirror symmetry of the
mesh. We furthermore reduce the influence of boundary effects on the
channel walls by prescribing periodic boundary conditions on both side
walls and the top and bottom of the channel. We run all simulations on the
larger time interval $[0.0,1.5]$. We have verified that all simulations are
past the transient startup and have reached a fully developed flow regime
at $t=0.5$. We take statistical averages on the time interval $[0.5,1.5]$.
The results are reported in Table~\ref{table:parameter_3d_mesh_distortion}.
It is striking that even a miniscule amount of \emph{nonuniform} mesh
distortion is sufficient to fully trigger the three dimensional instability
leading to streamwise vorticity. In particular there exists a bifurcation
point between $2.5\%$ mesh distortion and the fully symmetric mesh with
$0\%$ distortion at which the flow becomes potential;
cf.~Tables~\ref{table:parameter_3d},
and~\ref{table:parameter_3d_mesh_distortion}. However, we again observe a
strong dependence of the quantitative values on the level of mesh
refinement controlled by the refinement parameter $r$.

For completeness, a series of temporal snapshots of the simulations are
shown in Figure~\ref{fig:3d_iso_surfaces_mesh_distortion}, where,
consistent with previous visualizations, 30 equidistant pressure
iso-surfaces are plotted between the minimum and maximum pressure values.
We verify that the simulations do indeed reproduce the third observation
from \citep[p.~88]{Hoffman_2007}, namely, the strong generation of
streamwise vorticity. Our numerical findings thus support the conclusion
that the addition of nonuniform discretization errors through mesh
distortion destabilizes an otherwise potential-like flow, leading to
significant drag production~\citep[p.~88]{Hoffman_2007}. We
note however, that we are able again to tune quantitative results by about
an order of magnitude by controlling the refinement level $r$.


\section{Discussion and conclusion}
\label{sec:discussion}
The objective of this work was to investigate the influence of numerical
discretization errors on incompressible flow past a cylinder in two and
three spatial dimensions when equipped with the slip boundary condition.
The two main sources of discretization errors we examined were: (a) the
influence of the geometry approximation error at the curved boundary; and
(b) the influence of nonuniform artificial viscosity. The finite element
discretization of the Navier–Stokes equations was stabilized using a
Galerkin–Least-Squares (GLS) formulation.

To first minimize the influence of such discretization errors we employed
carefully constructed quasi-uniform quadrilateral and hexahedral meshes,
combined with a high-order isoparametric mapping to accurately represent
the cylinder boundary. Through a series of systematic numerical validations
in both two and three dimensions, we found that using a slip boundary
condition leads to a \emph{stable potential flow profile}, provided that
numerical surface roughness and mesh distortion are minimized. However, the
introduction of numerical discretization errors---whether through
inaccurate geometry representation or mesh distortion---destabilizes the
flow, resulting in nonzero drag and lift coefficients. Most importantly, we
have been able to demonstrate that numerical surface roughness and mesh
distortion can be used as a control parameter in this regard: By carefully
choosing the degree of boundary approximation error (or mesh distortion)
one can manipulate the drag and lift forces over a large interval of
numerical values spanning 3 to 5 orders of magnitude.

As a second numerical test we analyzed numerically the stability of the
potential flow simulation under random, nonuniform mesh distortion while
minimizing the influence of discretization errors on the boundary (by using
an isoparametric mapping and periodic wall boundary conditions for the
channel). It is striking that even a miniscule amount of nonuniform mesh
distortion is sufficient to fully trigger the three dimensional instability
leading to streamwise vorticity. We observe in particular that there exists
a bifurcation point between $2.5\%$ mesh distortion and the fully symmetric
mesh with $0\%$ distortion at which the flow becomes potential. However, we
again note that we observe a strong dependence of the quantitative values
on the level of mesh refinement.

Based on these findings, we conclude that slip boundary conditions---
at least in the form suggested by \citep{Hoffman_2007} with an under
resolved, coarse mesh and using GLS stabilization--- are not suitable for
\emph{wall-modeled} large-eddy simulation of realistic fluid flows at low
viscosity. Slip boundary conditions fail to accurately predict key
aerodynamic quantities such as drag and lift, and do not correctly capture
flow separation behavior.


\section*{Acknowledgments}

This work has been partially supported by the National Science Foundation
under grant DMS-2045636 (MM), by the Air Force Office of Scientific
Research, USAF, under grant/contract number FA9550-23-1-0007 (MM), and
Swedish Research Council under grant number 2021-04620 (PM \& MN). The
authors acknowledge fruitful discussions and collaborations with Bruno
Blais and Laura Prieto Saavedra as well as the larger \texttt{deal.II}
community.


\bibliographystyle{abbrvnat} 
\bibliography{ref}
\end{document}